\documentclass[%
twocolumn,
groupedaddress,
amsmath,amssymb,
aps,
pra,
]{revtex4-2}

\usepackage{graphicx}
\usepackage{amsfonts}
\usepackage{bm}
\usepackage{hyperref,url}
\usepackage{changes}
\usepackage{xcolor}
\usepackage{soul}

\newcommand{\bd}{\begin{displaymath}}
\newcommand{\ed}{\end{displaymath}}
\newcommand{\be}{\begin{equation}}
\newcommand{\ee}{\end{equation}}
\newcommand{\bs}{\begin{subequations}}
\newcommand{\es}{\end{subequations}}
\newcommand{\ba}{\begin{eqnarray}}
\newcommand{\ea}{\end{eqnarray}}

\begin{document}

\title{Decoherence in quantum cavities: Environmental erasure of carpet-type structures}

\author{E. Honrubia}
\email{efrenhon@ucm.es}
\affiliation{Department of Optics, Faculty of Physical Sciences,
Universidad Complutense de Madrid\\
Pza.\ Ciencias 1, Ciudad Universitaria -- 28040 Madrid, Spain}

\author{A. S. Sanz}
\email{a.s.sanz@fis.ucm.es}
\affiliation{Department of Optics, Faculty of Physical Sciences,
Universidad Complutense de Madrid\\
Pza.\ Ciencias 1, Ciudad Universitaria -- 28040 Madrid, Spain}

\date{\today}

\begin{abstract}
The interaction with an environment provokes decoherence in quantum systems, which gradually
suppresses their capability to display interference traits.
Hence carpet-type structures, which arise after the release of a localized state inside a
quantum cavity, constitute an ideal laboratory to study and analyze the robustness of the
interference process that underlies this phenomenon against the harmful effects of
decoherence.
Such a released localized state may represent a radiation mode inserted into a multimode
interference device or a cold-atom system released in an optical trap, for instance.
Here, without losing any generality, for simplicity, the case of a particle
with a mass $m$ is considered and described by a localized state corresponding to the ground state of a
square box of width $w$, which is released inside a wider cavity (with a width $L > w$).
The effects of decoherence are then numerically investigated by means of a simple
dynamical model that captures the essential features of the phenomenon under Markovian
conditions, leaving aside extra complications associated with a more detailed dynamical
description of the system-environment interaction.
As it is shown, this model takes into account and reproduces the fact that decoherence
effects are stronger as energy levels become more separated (in energy), which translates
into a progressive collapse of the energy density matrix to its main diagonal.
However, because energy dissipation is not considered, an analogous behavior is not observed
in the position representation, where a proper spatial localization of the probability density
does not take place, but rather a delocalized distribution.
This result emphasizes the fact that classicality is reached only if both decoherence and
dissipation coexist; otherwise, non-classical traits might still persist.
Actually, as it is also shown, in the position representation some off-diagonal correlations
indeed survive unless an additional spatial-type factor is included in the model.
This makes evident the rather complex nature of the decoherence phenomenon and hence the
importance to have a familiarity with how it manifests in different representations,
particularly with the purpose to determine and design reliable control mechanisms.
\end{abstract}

\maketitle


\section{\label{sec1} Introduction}

Decoherence is central to the description and understanding of quantum systems whenever
they are not under ideal isolation conditions \cite{breuer-bk:2002}.
Their interaction with other surrounding systems leads them to gradually lose their
coherence properties and hence to exhibit behaviors that resemble those typical of classical
systems.
This effect has been commonly referred to in the literature as the emergence of the classical
world \cite{joos-zeh:ZPhysB:1985,zurek:PhysToday:1991,zurek:PhysToday-rev:2003,blanchardolkiewicz,giulini-bk,schlosshauer:RMP:2004,schlosshauer-bk:2007}.
Decoherence is ubiquitous in a myriad of systems and applications \cite{schlosshauerdecoh},
e.g., quantum dots \cite{khaetskii:PRL:2002,mathew:MPLB:2013,altaisky:EPJWebConf:2016}, quantum game theory
\cite{flitneyabbot,prisoner}, quantum walks \cite{kendon:PRA:2003,yin:PRA:2008}, quantum information
\cite{milburn1,leman:CTP:1998,kimm:PRA:2002,wudenglidassarma}, two-level systems
\cite{ziman:PRA:2005,czerwinski:IJTP:2016,floss:PRB:2019}, cavities \cite{longhi,chen:CTP:2013,petruccione:PRAppl:2018}, ion
trapping \cite{schroll:PRA:2003} or the spin-boson model \cite{dehdashti,schlosshauerdecoh}.
Similarly different models have been proposed to study and quantify its effects
on the coherence of quantum systems as well as to control them
\cite{duan:ChinPhysLett:1997,duanguo2,duanguo4,shor,violalloyd,novais:PRL2006,alicki,lei:PRA:2011,kimlee,bi,ahsan:QIC:2018}.
Furthermore, a number of experiments have also been conduced in recent year to test such models at a fundamental level \cite{ralphpienaar,batelaan:NJP:2018,zeilinger:NJP:2018,xu}.

Depending on the system or, to be more precise, the nature and strength of its interaction
with a surrounding environment (many times also on the intrinsic properties of this
environment), different alternative theoretical models can be used to model the effects
of decoherence \cite{breuer-bk:2002,schlosshauerdecoh}.
These models may consist of simple sets of differential equations in terms of energy levels,
such as the Bloch equations, or more general equations of motion, such as the Lindblad
equation, valid in any representation.
However, taking such theoretical models and their outcomes as the basis, it is still possible
to build simpler phenomenological models that capture the physical essence of the phenomenon
and, by adjusting a few parameters, also provide us with a clear picture of the processes
involved at different levels of detail.
This is of particular interest in the analysis of highly intricate structures generated
by interference, as it is the case of quantum carpets, which generate inside cavities by
linearly and coherently superimposing a number of energy eigenstates
\cite{kaplan:PhysScr:1998,marzoli:ActaPhysSlo:1998,kaplan:PRA:2000,berry:PhysWorld:2001}.
It is clear that, as the number of eigenstates increases and the frequencies involved in
the superposition become higher and higher, the system becomes more sensitive to decoherence.
Hence quantum carpets seem to constitute an ideal scenario to explore the effects of decoherence
\cite{bonifacio,schleich:NJP:2013}.

In this work we analyze the consequences of a purely decoherent model on the erasure of quantum
carpets in terms of the symmetries displayed by the latter.
More specifically, here we consider the carpets developed upon the release of an initially
localized matter waved describing a particle with mass $m$ for simplicity, although without
any loss of generality, since the treatment can equally be applied to light carpets inside
confining cavities (resonant cavities or multimode interference devices).
This analysis, performed in the position representation as well as in the energy representation,
renders an interesting picture on how decoherence acts in each case.
Eventually this serves to better understand more refined and exact models, where such splitting
is not possible due to their intrinsic formal and conceptual nature.
Under Markovian conditions decoherence has important effects on both the position and the energy
representations \cite{sanz:CJC:2014}.
Thus, the model considered here relies on these observations and consists of a simple Markovian
coherence-damping term where energy differences between eigenstates and two-point position
correlations appear separately.
As it is shown, this leads to a bare addition of level populations.
In the energy representation, this translates into a gradual suppression of the off-diagonal
elements of the density matrix, only surviving the elements of the main diagonal, i.e., the
elements that physically account for the populations.
In the position representation, on the other hand, it is observed that the probability density 
approaches a nonhomogeneous, delocalized density distribution along the cavity.
Furthermore, in terms of the density matrix in the position representation, it is also noticed
that the probability accumulates not only along the main diagonal, but also
along the secondary diagonal when the initial state displays an even symmetry with respect
to the center of the cavity.
In order to remove such a nonphysical behavior, associated with the symmetry of the system,
an additional decoherence term depending on two-point correlations has to be explicitly taken
into account, which explains the typical exponential decays in terms of factors of the form
$(x-x')^2$ that appear in spatial decoherence models \cite{giulini-bk}.
Furthermore, in order to provide a clearer picture of the decoherence dynamics, i.e.,
the transition from a highly organized interference-mediated structure to a stationary
(equilibrium) state due to decoherence, the generation of the quantum carpet has also been
monitored with the aid of Bohmian trajectories, which have already been used to explore
analogous effects in the context of the two-slit experiment
\cite{sanz:EPJD:2007,sanz:CPL:2009-2,luis:AOP:2015}.

The work is organized as follows.
Section~\ref{sec2} deals with the theoretical treatment of the time evolution of localized
states or signals freely released in the cavity and the subsequent emergence of quantum
carpets.
It also includes a brief overview of the Bohmian-type methodology that will be used to visualize
and hence to better understand the evolution of the probability distribution in terms of density
streamlines or Bohmian trajectories \cite{sanz:JPA:2008,sanz:FrontPhys:2019}.
Furthermore, several cases of symmetric and asymmetric carpets are presented and discussed with
the purpose to serve later on to evaluate the effects of decoherence.
In particular, the carpets considered arise from the time-evolution of single (symmetrically and
asymmetrically) localized signals and coherent superpositions made of two initially (and
symmetrically) localized signals.
In Sec.~\ref{sec3} the decoherence model is introduced from the standard point of view and also
within the Bohmian context;
the effects of this model on quantum carpets are
then analyzed and discussed in both the position representation and the energy representations.
We conclude with a summary and discussion in Sec.~\ref{sec4}.


\section{\label{sec2} Quantum carpet dynamics}


\subsection{\label{sec21} General aspects}

Quantum carpets, the highly symmetric pattern displayed in both space and time by
the probability density inside a cavity, arise as a consequence of a rather complex
interference process involving a number of energy eigenstates (vibrational modes) of
the cavity \cite{berry:PhysWorld:2001}.
This is the behavior, for instance, that follows after pumping a localized state into a cavity
and then letting it freely evolve inside such a cavity.
The state can be a mode propagated along an optical fiber and then released into a broader
cavity, such as a multimode interference device \cite{newref1,pennings:JLightTech:1995,newref3}, or a
cold-atom system confined and guided in an optical trap \cite{newref4}.
In either case, as soon as the localized wave describing the state of the system is released
inside the wider cavity, it starts reconfiguring according to the new boundaries, thus giving
rise to the appearance of interference traits, which depend on the shape of the initially
localized state and the size of the cavity \cite{sanz:PhysScr:2019}.
To better understand the process and also to be self-contained, let us start by briefly
describing the process that leads to the appearance of quantum carpets as well as some
of its most relevant properties in connection to this work.

Thus, consider that the matter wave associated with a particle with mass $m$ is pumped into
a cavity of a certain width, which will be assumed to be one dimensional for simplicity.
Such a matter wave is describable in terms of a localized wave function, which will be
referred to from now on as the input signal, making use of a more operational description, also
valid in the optical scenario \cite{footnote}.
Note that this input signal describes the particular shape of the input beam, which can be
a single vibrational mode or eigenstate transported from a narrower cavity (waveguide) to
the new one.
This is the case here considered, where the, say, collecting cavity is assumed to be a
square box centered at $x=0$ and with a width $L$, larger than the typical width associated
with the input signal, henceforth denoted by $w$.

As it is well known, in such a case, the input signal can then easily be recast as a
coherent superposition of the corresponding basis set of energy modes (eigenstates)
$\{\varphi_\alpha(x) \in \mathbb{R}, \alpha = 0, 1, 2, \ldots, |x| \le L/2\}$ as
\be
 \psi_0(x) = \sum_\alpha c_\alpha \varphi_\alpha (x) .
 \label{eq:ondainicial}
\ee
Each coefficient $c_\alpha$ is obtained from the projection of the input signal
onto the corresponding mode, i.e.,
\be
 c_\alpha = \int \varphi_\alpha^* (x) \psi_0 (x) dx .
\ee
If these coefficients are recast in polar form, i.e., as
$c_\alpha = |c_\alpha| e^{i\delta_\alpha}$,
it is clear that they contain information about how much each mode
contributes to the superposition, with the weight of such a contribution given by 
$|c_\alpha|^2$.
Hence, from now on, because they indicate the population of each mode, we will refer
to them as the populations, also in compliance with the convention commonly used in
level systems.
In the same way, the crossed terms, $c_\alpha c_{\alpha'}^*$, will be referred to as
coherences, since they carry information about the mutual correlation between different
pairs of modes.
Since each coefficient carries a time-independent relative phase $\delta_\alpha$, they
will contribute to introduce relative phase differences among different modes, thus
generating constructive or destructive interference among them.
As for the modes, they are simple sinusoidal functions with even and odd parity
\cite{schiff-bk}
\begin{subequations}
\ba
 \varphi_\alpha^e(x) & = & \sqrt{\dfrac{2}{L}}\cos (k_\alpha x) , \\
 \varphi_\alpha^o (x) & = & \sqrt{\dfrac{2}{L}}\sin (k_\alpha x) ,
\ea
 \label{eq:eigenfunc}
\end{subequations}
with $k_\alpha = \alpha\pi/L$, where $\alpha = 2n-1$ for the even-parity solutions ($e$)
and $\alpha = 2n$ for the odd-parity solutions ($o$), with $n=1,2,\ldots$ in both cases.
The corresponding eigenenergies are $E_\alpha = \hbar^2 k_\alpha^2/2m = \hbar^2 \pi^2 \alpha^2/2mL^2$.

The spectral decomposition accounted for by Eq.~(\ref{eq:ondainicial}) enables a simple
description of the evolution of the input signal at any subsequent time.
As it is well known, because each mode is associated with a specific energy $E_\alpha$,
the time evolution of Eq.~(\ref{eq:ondainicial}) is analytical and has the simple functional
form
\be
 \psi(x,t) = \sum_\alpha c_\alpha \varphi_\alpha (x) e^{-i E_\alpha t/\hbar} .
 \label{eq:ondatiempo}
\ee
Due to the extra complex factors $\exp(-iE_\alpha t/\hbar)$, as soon as $t$ changes, all
spectral components start vibrating, thus changing their local value, which translates in
the aforementioned complex interference process, with probability distributions varying
from time to time.
However, although at some times the probability density might not keep any resemblance with
the original one, at other particular times it is possible to observe the appearance of
recurrences (a number of identical copies of the initial probability density distributed 
across the cavity) and revivals (a full reconstruction of the initial probability density).

The emergence of recurrences and revivals is better appreciated by explicitly computing the
probability density associated with (\ref{eq:ondatiempo}),
\ba
 \rho(x,t) & = & \sum_\alpha |c_\alpha|^2 \varphi_\alpha^2 (x) \nonumber \\
 & + & \sum_{\alpha' > \alpha} |c_\alpha| |c_{\alpha'}| \varphi_\alpha (x) \varphi_{\alpha'} (x) \cos (\omega_{\alpha \alpha'} t - \delta_{\alpha \alpha'}) ,
 \nonumber \\ & &
 \label{interf}
\ea
where
\be
 \omega_{\alpha \alpha'} \equiv  \frac{E_{\alpha'} - E_\alpha}{\hbar}
  = 2\pi \left( \frac{\pi \hbar}{4mL^2} \right) \left( \alpha'^2 - \alpha^2 \right)
 \label{freqdef}
\ee
and $\delta_{\alpha \alpha'} \equiv \delta_{\alpha'} - \delta_\alpha$, with $\alpha' > \alpha$.
Note that the splitting in (\ref{interf}) makes explicit the contribution arising from
populations and coherences independently.
As it can be seen, this splitting is rather convenient, because all the dynamics is
contained in the coherences; populations remain unchanged unless dissipation is present,
which is not the case here.
Our decoherence model will focus on the second term of Eq.~(\ref{interf}), which
defines the phase relations between the different modes through the relative phase shifts
$\delta_{\alpha \alpha'}$.
In the latter regard, since the input signal has no transverse displacement component,
we have $\delta_{\alpha \alpha'} = 0$ for all $\alpha$ and $\alpha'$.
Those displacements imply the presence of extra factors, of the kind
$e^{\pm i k x}$ in $\psi_0(x)$, which eventually turn into nonvanishing
relative phase shifts.
However, this will not be the case here, where input signals are assumed to be pumped
into the cavity without lateral motion.

In spite of the complex interference process described by (\ref{interf}) and
the shape displayed by the input signal, it is easy to see that there is a
revival of the state after some time.
More specifically, this happens whenever time is such that the time-dependent
phase factor $\omega_{\alpha \alpha'} t$ is an integer multiple of $2\pi$ for
all $\alpha$ and $\alpha'$.
If we denote by $T_{\rm rev}$ the first time $t$ at which this condition is satisfied,
then we find that
\be
 \omega_{\alpha \alpha'} T_{\rm rev} = 2\pi \left( \frac{\pi \hbar}{4mL^2} \right)
  \left( \alpha'^2 - \alpha^2 \right) T_{\rm rev} .
\label{eq:periodo}
\ee
Because $\alpha'^2 - \alpha^2$ is a positive integer, the condition to
observe the first revival requires that
\be
 T_{\rm rev} = \frac{4mL^2}{\pi\hbar} .
 \label{rev}
\ee
This condition is of general validity regardless of the initial state $\psi_0(x)$ and
its spectral decomposition.
Accordingly, whenever $t = \nu T_{\rm rev}$, with $\nu = 1, 2, \ldots$, the probability
density undergoes a full revival, which implies that the signal $\psi(x,t)$ looks
the same as $\psi_0(x)$ in amplitude, but is affected by a global phase factor $2\pi\nu$.
Nonetheless, it is worth noting that, when the $2\pi\nu$ constraint is relaxed to
$\pi(2\nu-1)$, the mirror-symmetric (with respect to $x=0$) replica
of the initial state (discussed below) is observed, which cannot be considered as a proper recurrence.
Furthermore, if the input signal has a definite even or odd symmetry (i.e., it
consists of a superposition of only even or odd eigenfunctions, respectively), revivals
actually take place in shorter times, which does not invalidate the above condition.
Note that in both cases the factor $\alpha'^2 - \alpha^2$ in Eq.~(\ref{eq:periodo}) is
proportional to 4.
Indeed, because now all spectral components require the parity of the signal
to be in phase, the constraint to $2\pi\nu$ can be relaxed to simply $\pi\nu$.
Hence, the observation of revivals occurs at an eighth of the general revival time
$T_{\rm rev}$.
We define this new timescale as
\be
 \tau = \frac{T_{\rm rev}}{8} = \frac{mL^2}{2\pi\hbar} ,
 \label{min}
\ee
which will be used from now on as the reference time.

The presence of revivals of the initial state as well as recurrences at fractional times
gives rise to a pattern with space and time symmetries referred to as a quantum carpet
\cite{berry:PhysWorld:2001}.
Typically the concept of quantum carpet is associated with the probability density, although
similar symmetries can also be found in the amplitude and phase of the signal, or the
quantum flux, which becomes more apparent when it is analyzed in terms of the associated
streamlines or Bohmian trajectories \cite{sanz:PhysScr:2019}.
In order to determine the effects of decoherence on the flow of probability inside the
cavity, here we proceed similarly, i.e., considering the supplementary aid of a
Bohmian-type description.
In brief, this formulation readily arises after recasting the signal in polar form
\cite{sanz:FrontPhys:2019}, which provides us with a nonlinear transformation from a complex
field to two real-valued ones, namely, the probability density $\rho$ and the phase $S$,
\be
 \psi (x,t) = \rho^{1/2} (x,t) e^{iS(x,t)/\hbar} .
 \label{eq:dpfase}
\ee
It is well known \cite{schiff-bk} that, by virtue of the continuity equation for the
probability density, it is possible to establish a relation between the latter and its
flux through an underlying velocity field.
In one dimension, this reads
\ba
 J(x,t) & = & \frac{\hbar}{2mi} \left[ \psi^*(x,t) \partial_x \psi(x,t)
  - \psi(x,t) \partial_x \psi^*(x,t) \right] \nonumber \\
 & = & v(x,t) \rho(x,t) ,
 \label{quantumflux}
\ea
where the shorthand notation $\partial_x \equiv \partial/\partial x$ is used for
convenience.
Following this relation, now we have an unambiguous mechanism to monitor at a local
level, i.e., at each point, how the probability density contracts and expands by
interference as it flows across it back and forth inside the cavity.
More specifically, these variations take place in compliance with the underlying
local variations of the velocity field $v(x,t)$,
which depend in turn on the local variations undergone by the phase field:
\be
 v(x,t) = \frac{J(x,t)}{\rho(x,t)} = \frac{\partial_x S(x,t)}{m} .
 \label{velocity}
\ee
Of course, if there is a local velocity field, it is straightforward that, given any
position, a trajectory can be obtained by integrating this coordinate-dependent and
time-varying field in time, i.e., by integrating the equation of motion
\ba
 \dot{x} & = & v(x,t) \nonumber \\
 & = & \frac{\hbar}{2mi}
  \left[ \frac{\psi^*(x,t) \partial_x \psi(x,t)
  - \psi(x,t) \partial_x \psi^*(x,t)}{\psi^*(x,t) \psi(x,t)} \right] .
 \label{eom}
\ea
These trajectories are the so-called Bohmian trajectories in the literature \cite{holland-bk}
and correspond to streamlines along which probability flows.
Here they will provide us with a better idea of how the gradual suppressing effects led by
decoherence are going to affect the degradation of quantum carpets, since Eq.~(\ref{eom})
contains relevant information about the coherences.
This is readily seen by substituting Eq.~(\ref{eq:ondatiempo}) into (\ref{eom}), which
gives the analytical functional form
\begin{widetext}
\be
 \dot{x} = \frac{\hbar}{m}
  \left\{ \frac{\sum_{\alpha' > \alpha} |c_\alpha| |c_{\alpha'}|
   \left[\varphi_{\alpha'} (x) \partial_x \varphi_\alpha (x)  - \varphi_\alpha (x) \partial_x \varphi_{\alpha'} (x) \right] \sin (\omega_{\alpha \alpha'} t - \delta_{\alpha \alpha'})}{\sum_\alpha |c_\alpha|^2 \varphi_\alpha^2 (x)
 + \sum_{\alpha' > \alpha} |c_\alpha| |c_{\alpha'}| \varphi_\alpha (x) \varphi_{\alpha'} (x) \cos (\omega_{\alpha \alpha'} t - \delta_{\alpha \alpha'})} \right\} ,
 \label{eom2}
\ee
\end{widetext}
which is very convenient both at a numerical level and also in order to determine the
effective action of decoherence (discussed below in Sec.~\ref{sec3}).

Note that the same procedure can also be straightforwardly applied to light carpets inside
optical fibers, resonant cavities, or multimode couplers \cite{sanz:JOSAA:2012}.
In such a case, the modes correspond to electromagnetic modes in compliance with the
solutions to the Helmholtz equation for a cavity and the flux is determined by the Poynting
vector (the flux of electromagnetic energy).
Actually, if the Helmholtz equation is in paraxial form, its solutions will be formally
equivalent (isomorphic) to those of the Schr\"odinger equation, except for the replacement
of time by the longitudinal coordinate \cite{sanz-bk-1}.


\subsection{\label{sec22} Symmetric and asymmetric carpets}

If the lowest (fundamental) mode transported by an optical fiber is pumped into a
wider cavity, a multimode interference device \cite{pennings:JLightTech:1995}, the light
distribution inside the new space will exhibit carpet-type traits \cite{berry:PhysWorld:2001},
since such a mode can now be described as a coherent superposition of proper modes of the
cavity.
To recreate this kind of behavior in a quantum context, here we consider single
half-cosine amplitudes and a coherent superposition of them as initial {\it Ans\"atze}.
The cavity is simulated by means of a one-dimensional infinite well with a width $L=50$
(arbitrary units), centered around $x=0$, while the width of the half-cosine input amplitude
is $w=10$.
The general expression for spectral decomposition of the input signal which will be
considered is of the type
\be
 \psi_0 (x) = \sum_{{\rm even}\ \alpha} c_\alpha^e \varphi_\alpha^e (x)
  +\sum_{{\rm odd}\ \alpha} c_\alpha^o \varphi_\alpha^o (x) ,
 \label{superp}
\ee
where $\varphi_\alpha^e$ and $\varphi_\alpha^o$ are as specified by (\ref{eq:eigenfunc}).

\begin{figure}[t]
 \centering
 \includegraphics[width=\columnwidth]{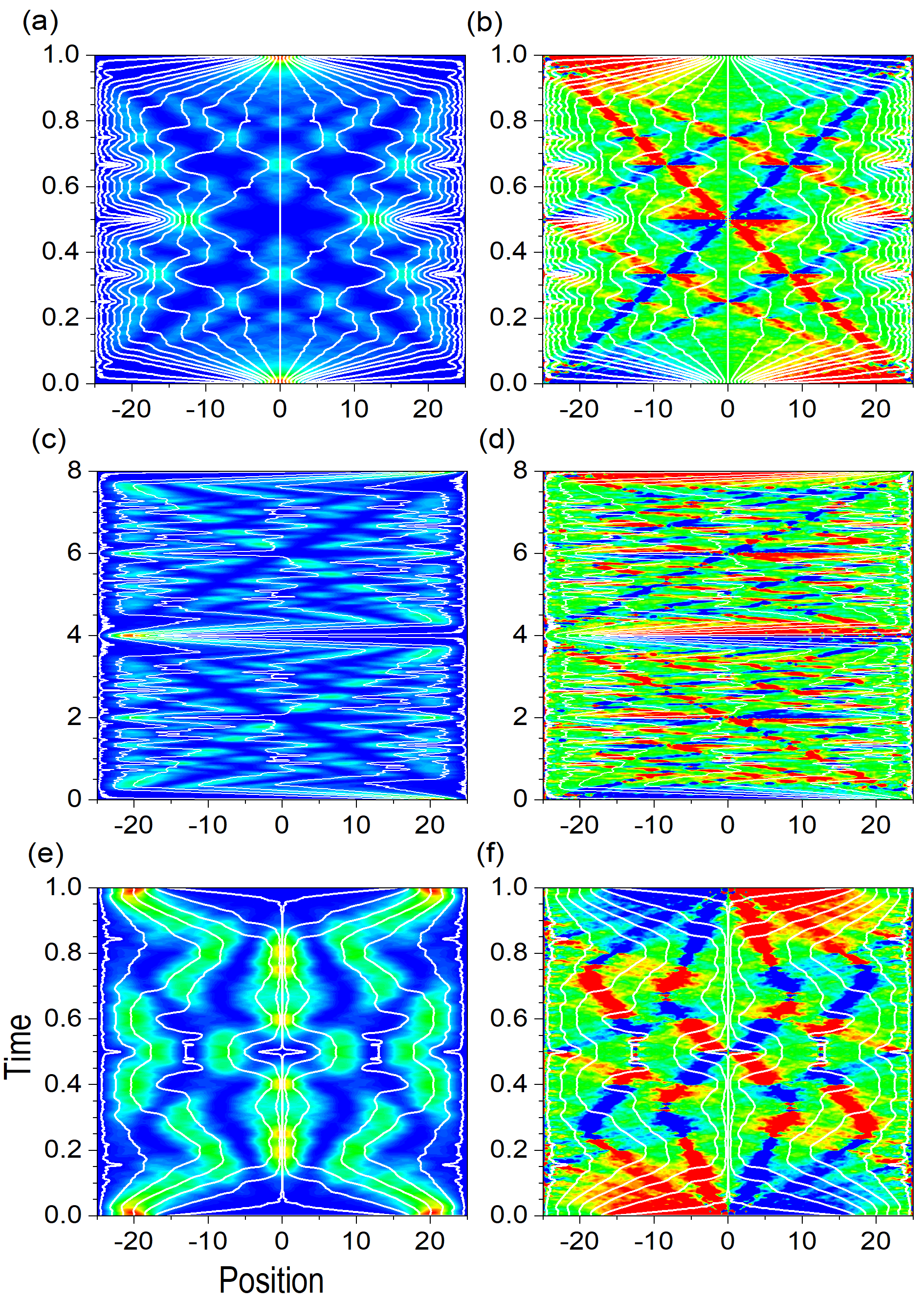}
 \caption{Quantum carpets generated by half-cosine-type amplitudes: probability density (left
  column) and velocity field (right column).
  In all cases, $L=50$, $w=10$, $m=1$ and $\hbar=1$ (all quantities are given in
  arbitrary units).
  Time is measured in units of $\tau$, as defined in Eq.~(\ref{min}).
  From top to bottom, single half-cosine input amplitudes centered at $x_0=0$ (upper row)
  and $x_0 = 20$ (middle row), and coherent superposition of two of them centered at
  $x_0 = \pm 20$ (lower row).
  In the color scale for the density plots, blue is used to denote the lowest values
  (zero values for the probability density and maximum negative values for the velocity
  field) and red the highest values (maximum positive values for the velocity field).
  To better specify the flow inside the cavity, a number of Bohmian trajectories (white
  solid lines) have been included in each panel.}
\label{fig1}
\end{figure}

In the case of a single half-cosine centered at $x=x_0$, the initial amplitude is given by
\be
 \psi_0(x) = \left\{ \begin{array}{ccc}
  \displaystyle \sqrt{\dfrac{2}{w}} \cos \left[ \frac{\pi (x-x_0)}{w} \right] ,
  & \quad & \displaystyle |x-x_0| \leq \frac{w}{2} ,
   \\ & \quad & \\ 0 , & \quad & {\rm otherwise} .
   \end{array}\right.
 \label{eq:coseq}
\ee
When recast in the form (\ref{superp}), the coefficients $c_\alpha$ read
\be
 c_\alpha^e = \left\{ \begin{array}{lc}
  \displaystyle \frac{4}{\sqrt{wL}} \left( \frac{k_0}{k_0^2 - k_\alpha^2} \right)
   \cos (k_\alpha x_0) \cos \left( \frac{k_\alpha w}{2} \right) , & k_\alpha \ne k_0 ,
   \\ & \\
  \displaystyle \sqrt{\frac{w}{L}} \cos (k_0 x_0) & k_\alpha = k_0 , \end{array} \right.
 \label{decomp1}
\ee
with $\alpha = 2n-1$, and
\be
 c_\alpha^o = \left\{ \begin{array}{lc}
  \displaystyle \frac{4}{\sqrt{wL}} \left( \frac{k_0}{k_0^2 - k_\alpha^2} \right)
   \sin (k_\alpha x_0) \cos \left( \frac{k_\alpha w}{2} \right) , & k_\alpha \ne k_0 ,
   \\ & \\
  \displaystyle \sqrt{\frac{w}{L}} \sin (k_0 x_0) & k_\alpha = k_0 , \end{array} \right.
 \label{decomp2}
\ee
with $\alpha = 2n$.
In both cases, $k_\alpha = \alpha\pi/L$ and $k_0 = \pi/w$ (notice that the condition
$k_\alpha = k_0$ simply means that $w$ is an integer, even or odd, fraction of $L$).
The quantum carpets generated by initial amplitudes centered at $x_0=0$ and $20$ are
displayed in Figs.~\ref{fig1}(a) and (c), respectively.
As it can be noticed, the loss of symmetry in the second case results in the revival
time being reached at $t=T_{\rm rev} = 8\tau$ instead of at $t=\tau$, which is the case in the
symmetric configuration.
Nonetheless, in both cases it is possible to observe the presence of fractional recurrences,
which is used in multimode interference devices (when dealing with light) to produce a
number of identical copies of the same input state.

In order to elucidate how the probability density evolves inside the cavity, i.e., how the
interference maxima that generate the recurrences arise, a number of Bohmian trajectories
have been considered, evenly distributing their initial positions along the extension of
the input signal.
It can be seen how the initial diffraction launches the trajectories in a relatively fast
manner towards the boundaries of the cavity, thus covering the whole available space inside
it.
Although the appearance of recurrences and revivals is independent of the input signal,
the initial boost strongly depends on it, as shown elsewhere \cite{sanz:PhysScr:2019}.
The same behavior can be observed in both symmetric and asymmetric cases, though with the
difference that in the latter case trajectories on one side of the input signal have to travel
a larger distance than those started on the other side.
Nonetheless, because of this asymmetry, the trajectories on the right side of $\psi_0$
gradually start being launched towards the left side of the cavity, until at
$t=T_{\rm rev}/2$ they all gather and produce a full revival of the input signal at
$x = -x_0 = -20$.
This intricate motion can be better understood by analyzing the carpets associated with the
velocity field, in Figs.~\ref{fig1}(b) and (d), respectively for each case.
As it can be noticed, the local velocity values reach very sudden changes (red and blue
regions), which act on the trajectories in the same way as bumpers and other targets in
a pinball machine.

So far, the presence of a single input signal only affects the interference
process that follows the diffraction of such a signal.
Another case of interest is that of two coherent input signals, analogous to a two-slit
experiment carried out inside the box.
In this regard, we consider the above asymmetric input and construct a
superposition with its mirror image, i.e., we now consider an input signal of the type
\be
 \Psi_0(x) = \left\{ \begin{array}{ccc}
  \displaystyle \sqrt{\dfrac{1}{w}} \cos \left[ \frac{\pi (x \pm x_0)}{w} \right] ,
  & \quad & \displaystyle |x \pm x_0| \leq \frac{w}{2} ,
  \\ & \quad & \\ 0 , & \quad & {\rm otherwise} , \end{array} \right.
 \label{eq:cosdobleeq}
\ee
with $x_0 = 20$.
Because of its even symmetry with respect to $x=0$, all odd-symmetric terms in
(\ref{superp}) vanish ($c_\alpha^o = 0$),
\be
 c_\alpha^e = \left\{ \begin{array}{lc}
  \displaystyle 4\sqrt{\frac{2}{wL}} \left( \frac{k_0}{k_0^2 - k_\alpha^2} \right)
   \cos (k_\alpha x_0) \cos \left( \frac{k_\alpha w}{2} \right) , & k_\alpha \ne k_0 ,
   \\ & \\
 \displaystyle \sqrt{\frac{2w}{L}} \cos (k_0 x_0) , & k_\alpha = k_0 , \end{array} \right.
 \label{decomp3}
\ee
with $\alpha = 2n-1$.
The corresponding carpets for the probability density and the velocity field are
displayed in Figs.~\ref{fig1}(e) and (f), respectively.
As it can readily be noticed, the presence of the second signal reduces the revival time
for the probability density to $\tau$, which does not depend on the particular choice of
$x_0$ but on the equal weight assigned to both signals.
Again, the pinball-type structure displayed by the velocity field becomes apparent, with
the addition of a sort of channeling pattern, which is a trait of an incipient Young-type
interference substructure \cite{tounli:arxiv:2021}, typical of two wave-packet superpositions
\cite{sanz:JPCM:2002,sanz:FrontPhys:2019}.


\section{\label{sec3} Decoherence-induced dynamics}


\subsection{\label{sec31} Effective modeling of decoherence}

Decoherence arises from the interaction of a quantum system with a surrounding environment
and manifests in the gradual loss of coherence, i.e., in its ability to display interference
\cite{giulini-bk,schlosshauer:RMP:2004}.
Formally, the different coupling of each system proper state with environmental states
eventually leads to the suppression of their interference due to the orthogonality of
the latter states \cite{omnes:RMP:1992}.
This is readily understood by inspecting its manifestation in the corresponding density
matrix elements.
Taking into account the states here, these elements are going to be of the form
\ba
 \rho(x,x';t) & = & \psi^*(x',t)\psi(x,t) \nonumber \\
 & = & \sum_{\alpha', \alpha} c_\alpha c_{\alpha'}^* \varphi_\alpha(x) \varphi_{\alpha'}(x') \nonumber \\
 & & \quad \quad \times
 e^{-i(E_\alpha - E_{\alpha'})t/\hbar} \mathcal{D}_{\alpha \alpha'}(x,x';t) ,
 \label{eq:ourmodel}
\ea
where $\mathcal{D}_{\alpha \alpha'}(x,x';t)$ accounts for the coherence loss term, which is
going to act on the off-diagonal elements of the density matrix.
Following models considered in the literature to describe two-state decoherence
\cite{joos:bk:1996,omnes:RMP:1992} as well as more detailed numerical models
\cite{brumer-elran:JCP:2004,brumer-elran:JCP:2013,sanz:CJC:2014}, clearly this term should
annihilate any correlation between two different energy states $\alpha$ and $\alpha'$ as well
as between two different spatial points $(x,x')$.
Furthermore, this coherence damping should show an exponentially decreasing behavior with
time.
Accordingly, we have considered a simple model for this dissipating term
\ba
 \mathcal{D}_{\alpha \alpha'}(x,x';t) = e^{- \beta_{\alpha \alpha'} t - \Lambda (x - x')^2 t} ,
 \label{eq:ourmodel-2}
\ea
where
\be
 \beta_{\alpha \alpha'} \equiv \gamma \omega_{\alpha \alpha'}
\ee
accounts for the gradual
suppression of the coherences between different modes, with $\gamma$ a coefficient used
for control (to speed up or to slow down the
decoherence rate, constant for all combinations of modes).
Note that since $\omega_{\alpha \alpha'}$ is positive [according to the
definition (\ref{freqdef})], $\beta_{\alpha \alpha'}$ is also positive,
thus ensuring a gradual damping with time.
On the other hand, the factor
\be
 \Lambda = \frac{2\pi\hbar}{m L^3}
 \label{localizrate}
\ee
is the usual localization rate \cite{joos-zeh:ZPhysB:1985,joos:bk:1996}, which determines the
how fast two points of the signal, at $x$ and $x'$, lose mutual coherence with time.
The particular choice here produces an overall decay going like $1/L$ for small
distances $|x-x'|$ (the energy term is the dominant one) and like $L$ for large ones (the
spatial term becomes the leading one, as required to annihilate the prevalence of coherence
along the secondary diagonal). 

By inspecting (\ref{eq:ourmodel}), we notice that the spatial part seems to play a
minor role when we look at the probability density, since this quantity only takes into
account the diagonal elements
\begin{widetext}
\be
 \rho(x;t) = \sum_\alpha |c_\alpha|^2 \varphi_\alpha^2 (x)
 + \sum_{\alpha' > \alpha} |c_\alpha| |c_{\alpha'}| \varphi_\alpha (x) \varphi_{\alpha'} (x) \cos (\omega_{\alpha \alpha'} t - \delta_{\alpha \alpha'})
   e^{- \beta_{\alpha \alpha'} t} ,
 \label{eq:ourmodel2}
\ee
which do not include nonlocal correlations but mutual coherence between energy proper
states, responsible for the appearance of the carpets.
Following the prescription provided in \cite{sanz:EPJD:2007}, the corresponding Bohmian-type
trajectories are given by the expression
\be
 \dot{x} = \frac{\hbar}{m}
  \left\{ \frac{\sum_{\alpha' > \alpha} |c_\alpha| |c_{\alpha'}|
   \left[\varphi_{\alpha'} (x) \partial_x \varphi_\alpha (x)  - \varphi_\alpha (x) \partial_x \varphi_{\alpha'} (x) \right] \sin (\omega_{\alpha \alpha'} t - \delta_{\alpha \alpha'})
   e^{- \beta_{\alpha \alpha'} t}}{\sum_\alpha |c_\alpha|^2 \varphi_\alpha^2 (x)
 + \sum_{\alpha' > \alpha} |c_\alpha| |c_{\alpha'}| \varphi_\alpha (x) \varphi_{\alpha'} (x) \cos (\omega_{\alpha \alpha'} t - \delta_{\alpha \alpha'})
   e^{- \beta_{\alpha \alpha'} t}} \right\} ,
 \label{eom2decoh}
\ee
\end{widetext}
which does not include any term depending on the spatial correlations, because the
evaluation is locally performed in contrast to two-slit scenarios, where the coherence loss
between spatially distinct states has to gradually vanish \cite{sanz:EPJD:2007}.
In spite of the apparently complex functional form displayed by Eq.~(\ref{eom2decoh}),
the asymptotic behavior of the trajectories can easily be inferred in the cases of strong
and weak decoherence.
Thus, for a strong decoherence (in terms of $\gamma$), in the leading term in the
denominator of Eq.~(\ref{eom2decoh}) is essentially dominated by the time-independent
population term, a rather weak time-dependent contribution still persists in the numerator
(in terms of the oscillatory functions, apart from the decaying factors).
This contribution is enough to make the trajectories approach their stationary state condition relatively fast.
This stationarity condition corresponds to the population distribution (discussed below).
On the other hand, for a weak decoherence, the rich structure displayed by trajectories
(see Fig.~\ref{fig1}) will get smoother until they completely stop and remain steady.
In either case, the trajectories will tend to distribute according to the bare sum of
partial distributions, the long-time limit of (\ref{eq:ourmodel2}), i.e.,
\be
 \rho_\infty (x) = \sum_\alpha |c_\alpha|^2 \varphi_\alpha^2(x) .
 \label{eq:densitydecoherence}
\ee
Accordingly, because there is no energy dissipation, there will not be spatial localization
either, which means that the trajectories will not remain within the region covered by
the input state.
On the contrary, they will distribute according (\ref{eq:densitydecoherence}) and satisfying
the so-called noncrossing rule, i.e., trajectories cannot cross one another, because nonlocal
spatial information still remains \cite{sanz:CPL:2009-2,luis:AOP:2015}.


\subsection{\label{sec32} Decoherence in the position representation}

\begin{figure}[!t]
 \centering
 \includegraphics[width=\columnwidth]{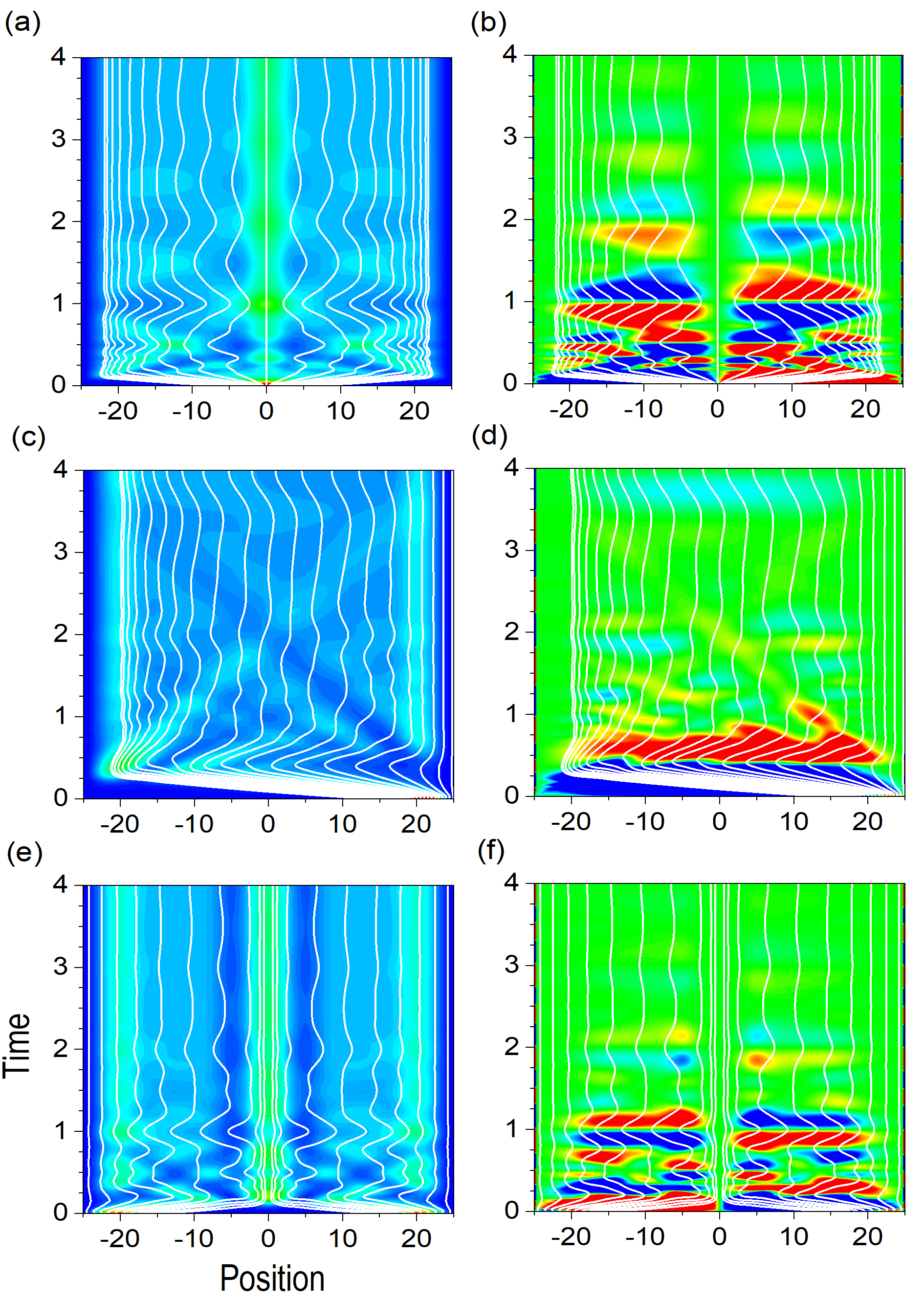}
 \caption{Same as in Fig.~\ref{fig1}, but with the presence of decoherence, with $\gamma=2/5\pi$.
  To better show the coherence loss effects, in all cases a total propagation
  up to four times the recurrence period of the symmetric case ($4\tau$) has
  been chosen.
  All quantities are given in arbitrary units.}
 \label{fig2}
\end{figure}

\begin{figure*}[t]
 \centering
 \includegraphics[width=\textwidth]{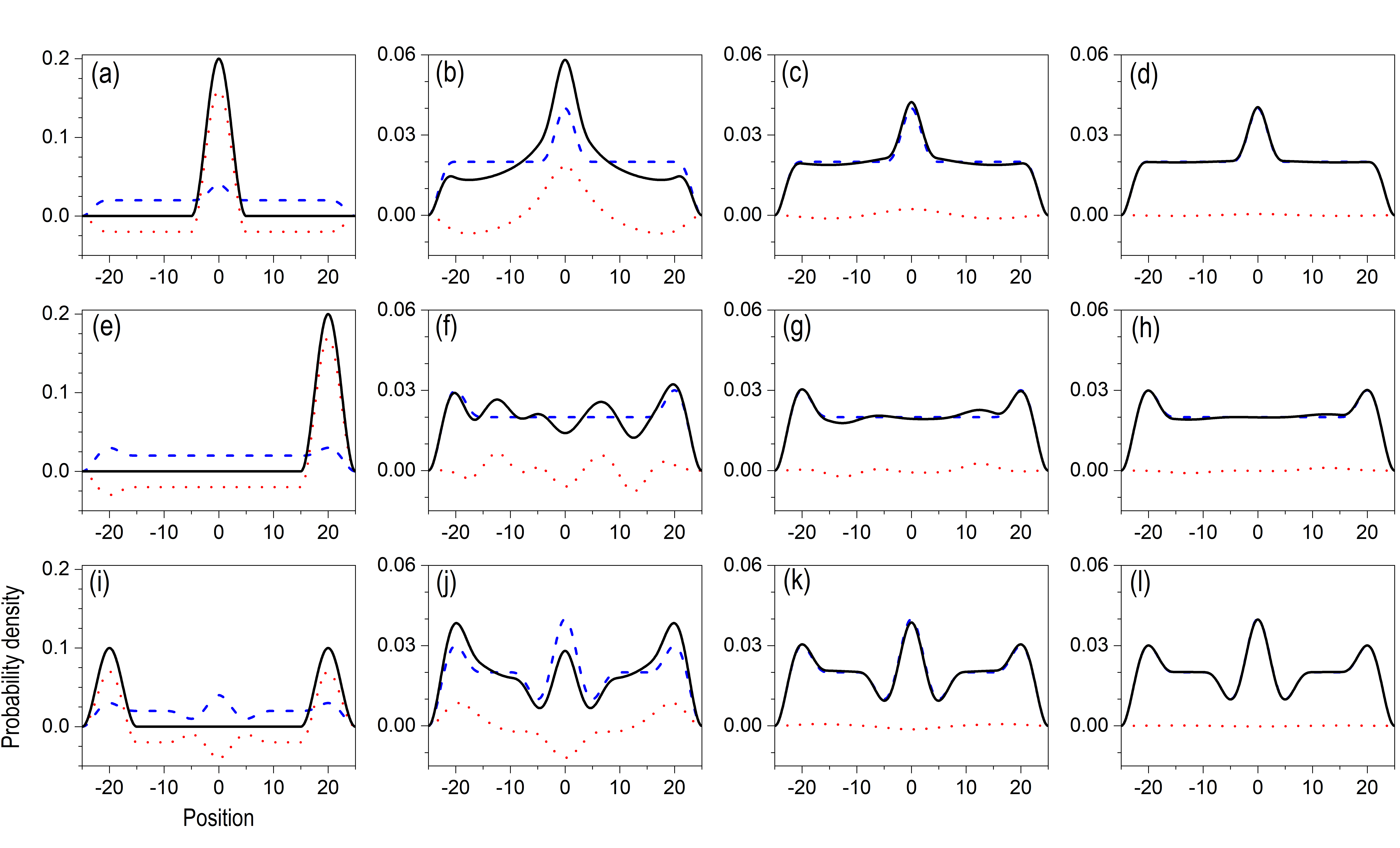}
 \caption{Snapshots of constant-time profiles for the three input states here considered
 (see Sec.~\ref{sec22}).
 From top to bottom: center-symmetric half-cosine ($x_0=0$), asymmetric half-cosine ($x_0=20$)
 and superposition of two half-cosines (with $x_0 = 20$).
 From left to right: $t=0$, $t=\tau$, $t=3\tau$ and $t=5\tau$.
 In each panel, the probability density is displayed with a black solid line, while the distributions
 related to the populations and the coherences are denoted by the blue dashed line and
 red dotted line, respectively.
 In all cases $L=50$, $w=10$, $m=1=\hbar$ and $N=50$.
 All quantities are given in arbitrary units.}
\label{fig3}
\end{figure*}

Taking into account the above-described effective model, let us know analyze the pattern
erasure undergone by the quantum carpets discussed in Sec.~\ref{sec22} in the presence of
decoherence.
To this end, for convenience but without any loss of generality, from now on we
consider $\gamma = 2/5\pi$.
With this value for the friction it is ensured that, at $t=\tau$, all energy-dependent decoherence factors amount to $({\alpha'}^2 - \alpha^2)/10$, thus giving rise to the
smoothing of the carpets in all cases in about four times $\tau$; of course, some terms
will decay faster than others depending on the difference $({\alpha'}^2 - \alpha^2)/10$.
Results for the three cases displayed in Fig.~\ref{fig1} are shown in the corresponding
panels of Fig.~\ref{fig2}.
As can be seen, the long-time limit in all cases consists of a series of steady trajectories,
unevenly but symmetrically distributed with respect to the center of the cavity, although one
might intuitively expect a relative localization around a certain position, in particular,
the center of the input distribution. 
This is in compliance, though, with the delocalization also exhibited by the probability density.
Note that in Fig.~\ref{fig2}(a), although a prominent central maximum becomes apparent as time
proceeds, we also observe that the probability distribution partly spreads out towards the
borders of the cavity.
This behavior finds a clear explanation when we look at the velocity field, shown in
Fig.~\ref{fig2}(b), which gradually loses its pinball-like structure and now presents an alternating distribution
of positive and negative regions.
As a consequence of this structure, the trajectories are smoothly driven until they completely
stop instead of undergoing sudden and fast motions.
To some extent, this situation is reminiscent of the calm waters that come after the rapids in a river,
which occurs when turbulence sources disappear.
In our context, this happens because the influence of the frequencies involved in the
interference process, which originates in nodal regions (vortices) and hence strong local
variations in the velocity field, is suppressed.
The same trend is observed in Fig.~\ref{fig2}(c).
Although the configuration is asymmetric,
it can also be seen that the trajectories smoothly move towards the symmetric position
(with respect to $x=0$) and then,
some of them, backward again, distributing nearly homogeneously across the cavity.
This is in compliance with the behavior shown by the velocity field in Fig.~\ref{fig2}(d).
Finally, the case of the superposition, shown in Fig.~\ref{fig2}(e), looks pretty similar
to that shown in Fig.~\ref{fig2}(a) for the symmetric single half-cosine, although there are two
additional accumulation regions at $x \approx \pm 20$, apart from the central one.

\begin{figure*}[t]
 \centering
 \includegraphics[width=\textwidth]{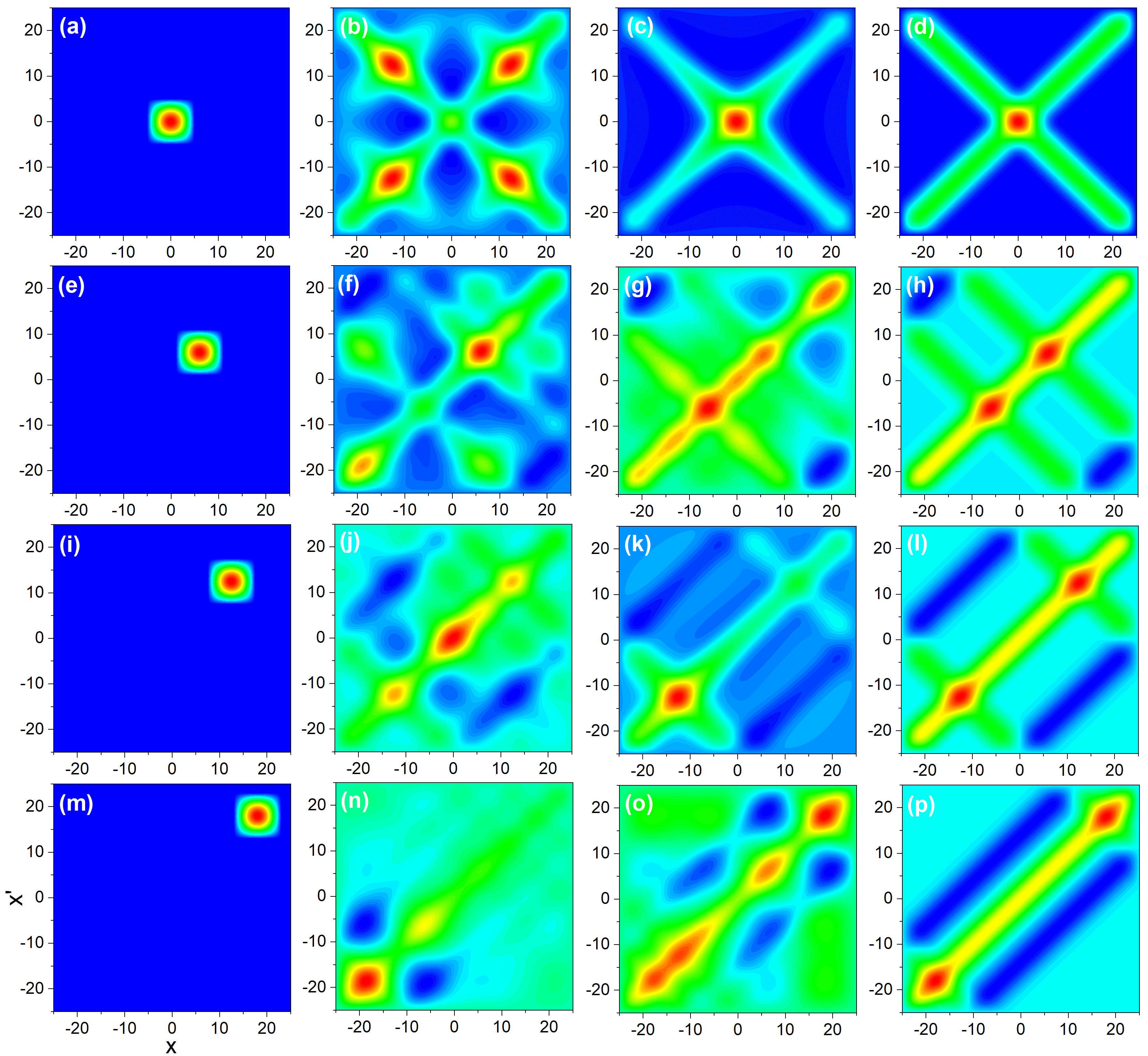}
 \caption{Real part of the density matrix for a single half-cosine initial input with
  $\Lambda=0$.
  From top to bottom rows: $x_0=0$, $x_0=6$, $x_0=12.5$ and $x_0=18$.
  From left to right: $t=0$,  $t=\tau/2$, $t=\tau$ and $t=20\tau$.
  As before, in all cases $L=50$, $w=10$, $m=1=\hbar$ and $\gamma=2/5\pi$.
  All quantities are given in arbitrary units.}
 \label{fig4}
\end{figure*}

\begin{figure*}[t]
 \centering
 \includegraphics[width=\textwidth]{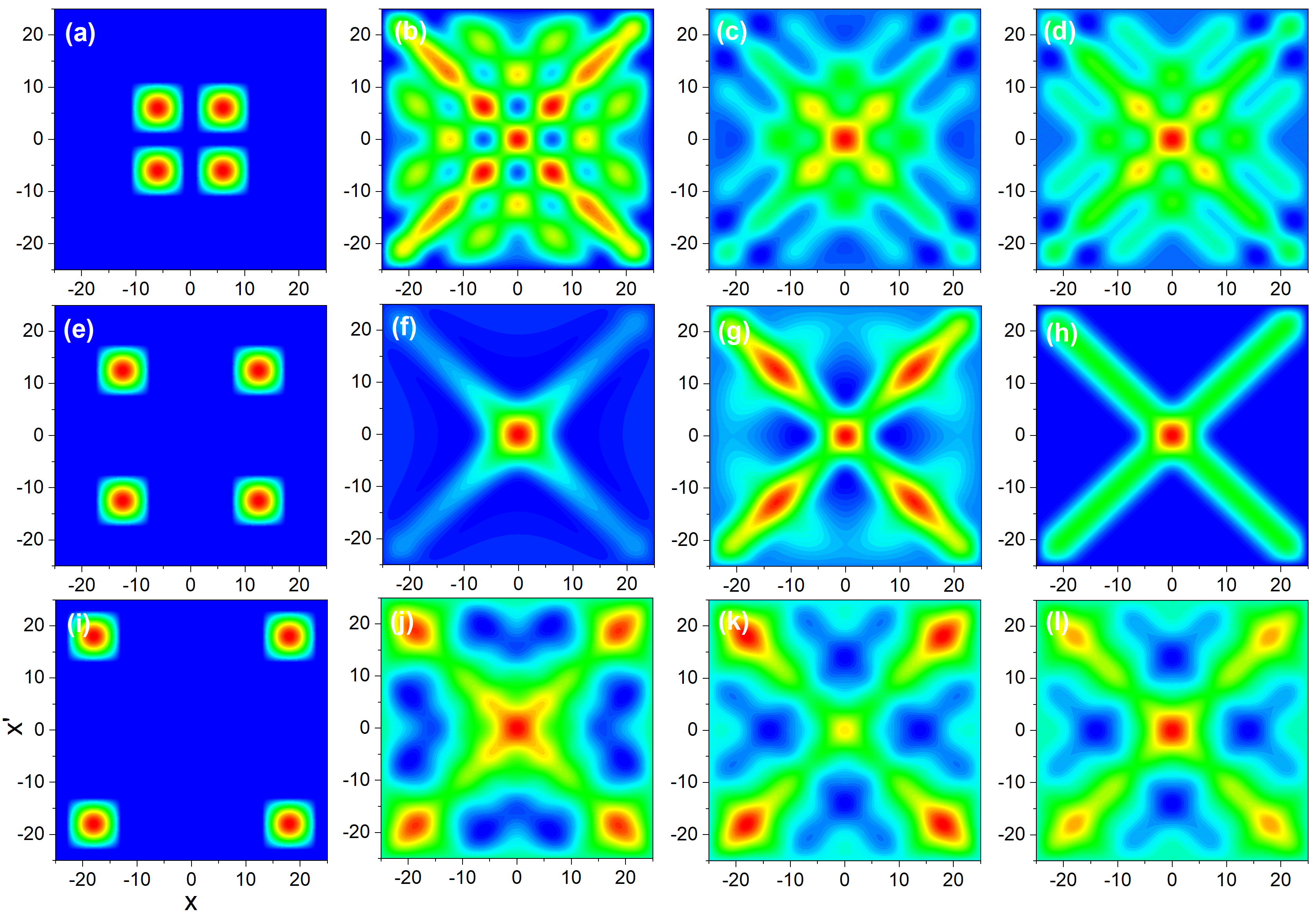}
 \caption{Real part of the density matrix for a coherent superposition of two half-cosine
  initial inputs with $\Lambda=0$.
  From top to bottom rows: $x_0=6$, $x_0=12.5$ and $x_0=18$.
  From left to right: $t=0$,  $t=\tau/2$, $t=\tau$ and $t=20\tau$.
  As before, in all cases $L=50$, $w=10$, $m=1=\hbar$ and $\gamma=2/5\pi$.
  All quantities are given in arbitrary units.}
 \label{fig5}
\end{figure*}

To better understand the behavior exhibited by both the probability density and the
trajectories in Fig.~\ref{fig2}, let us now analyze the time evolution of the probability
density, separating for our purpose the contributions coming from populations and
coherences.
Accordingly, some snapshots of the three quantities are displayed in Fig.~\ref{fig3} for
$t=0$, $\tau$, $3\tau$, and $5\tau$ (from left to right) for the three cases here considered
(from top to bottom).
As it can be noticed, at $t=0$ the probability density (black solid line) corresponds to
either single localized peaks or two peaks, depending on whether we have a single half-cosine
input signal or a superposition of two of them, respectively.
Now, if we inspect more closely these distributions, separating the contributions coming from
their populations (blue dashed line) and their coherences (red dotted line), some interesting
features readily emerge.
First of all, note that, as expected, the populations contribution is stationary, which is
due to the fact that this is a dissipation-free model, where thermalization effects that
reorganize populations in each cavity mode are disregarded.
As it was mentioned above, it is a purely decoherence model, which only influences the
state-state correlations.
Accordingly, although the initial probability density is peaked at some particular place or
places, the contribution associated with the populations exhibits a certain degree of delocalization across the cavity.
This is going to the asymptotic distribution once coherence is totally suppressed (see panels
in the last column).
Depending on whether a symmetric or an asymmetric input signal is considered, we observe
the presence or absence of a maximum at $x=0$, respectively.
This maximum plays the role of a certain effective barrier in the symmetric configurations:
Trajectories started on either side will never be able to cross to other side
\cite{sanz:JCP-Talbot:2007,sanz:PhysScr:2019}.
To some extent, the dynamics on either side of the central maximum is going to be ruled by
this maximum and the borders of the cavity, in a fashion similar to an effective
two-wave superposition (although with a more complex interference process in between, as
shown in Fig.~\ref{fig1}, top and bottom panels).
The same is not observed in the asymmetric case, because the absence of the central
maximum allows the trajectories to move from one side of the cavity to the other and vice
versa, since now these dynamics are ruled by the two marginal maxima, i.e., the one
corresponding to the input signal and its mirror image.
All these dynamics, on the other hand, are directly mediated by the changes in time undergone
by the coherence terms (red dotted line), i.e., the complex interference processes generated
by the correlations established among all modes.
When the decoherence damping factor starts acting on them, they gradually disappear, which
removes the oscillatory behaviors displayed with time and lets the so-far ``screened''
population contribution emerge, as it can be seen in the last two columns for all cases.
Accordingly, once the oscillatory (time-dependent) term has been canceled out, the
trajectories are  also going to stop their wandering motion inside the cavity, remaining
stationary, as seen in Fig.~\ref{fig2}.

\begin{figure*}[t]
 \centering
 \includegraphics[width=\textwidth]{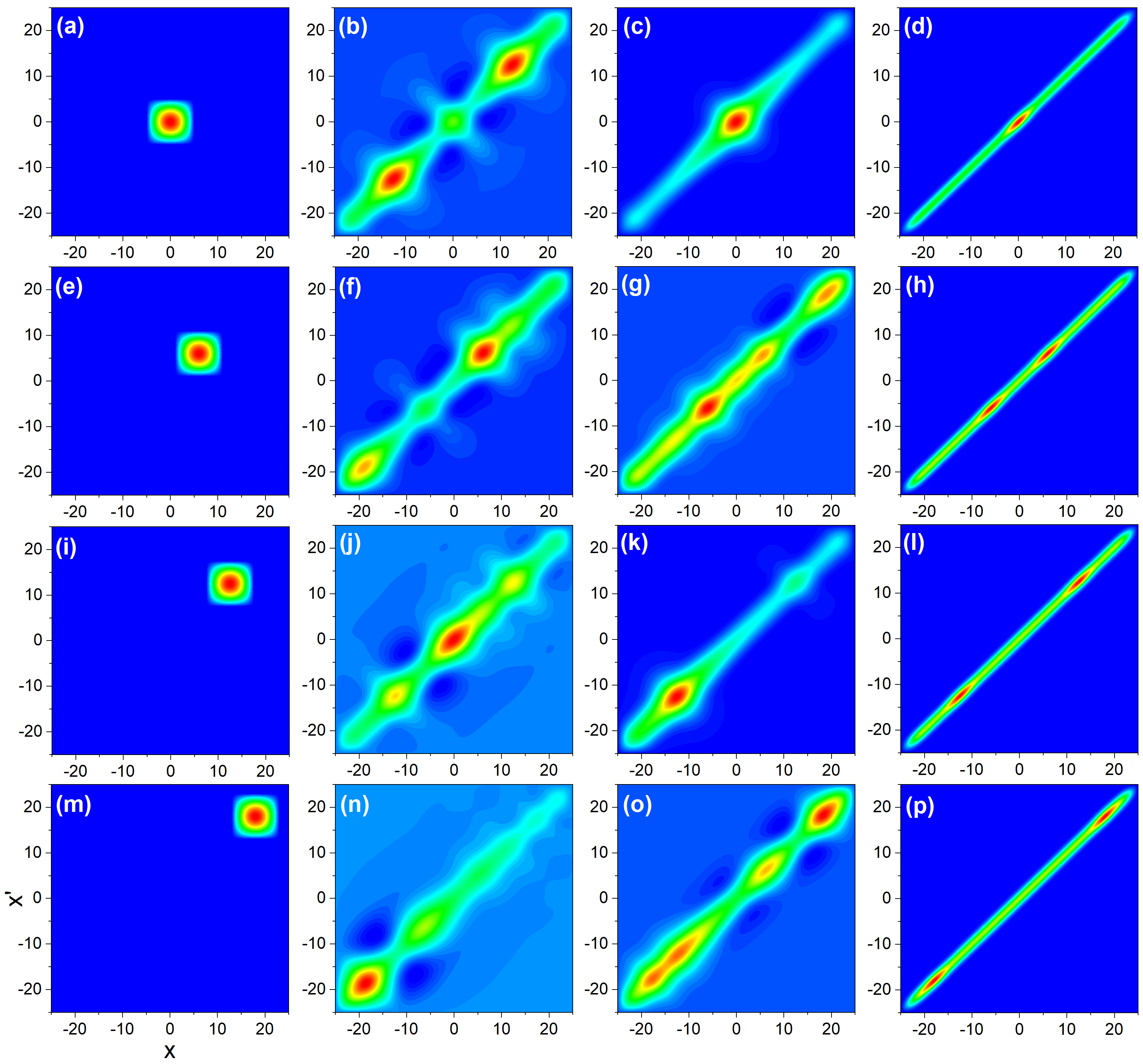}
 \caption{Same as in Fig.~\ref{fig4} but with $\Lambda$ as given by Eq.~(\ref{localizrate}).}
 \label{fig6}
\end{figure*}

So far we have focused on the probability density, trying to understand its dynamics in
terms of the underlying velocity field and the corresponding Bohmian trajectories.
Let us now consider the more general view provided by the corresponding density matrix,
firs assuming a vanishing localization rate, i.e., $\Lambda=0$, but keeping finite the
value of $\gamma$ (as before, $\gamma = 2/5\pi$).
When proceeding this way for a single input signal, we find that, as time proceeds, a rather
structured pattern arises independently of the value of $x_0$, as shown in Fig.~\ref{fig4}.
Each column shows a snapshot of the real part of the density matrix for the single
half-cosine initial {\it Ansatz}, namely, $t=0, \tau/2, \tau$, and $20\tau$, from
left to right, and four different positions of the signal center, namely, $x_0=0,
x_0=6, x_0=12.5$, and $18$, from top to bottom.
Intuitively, one would expect that, with time, the density distributes along the diagonal,
as it corresponds to the real part of the density operator, which what we are showing here.
However, what we observe is a very symmetric structure, which looks the same asymptotically along the main
diagonal and also on both sides with respect to the secondary diagonal
(last column).
This is because spatial correlations have not been properly removed, but only those in the
energy domain.
The same behavior is also observed in the case of the initial superposition, as it is
shown in Fig.~\ref{fig5} for $x_0 = 6, 12.5$, and $18$, from top
to bottom, and the same four times, where a rather four-fold symmetric pattern emerges.
Interestingly, the case for input signals at localized at the center of each
half of the cavity ($x_0 = 12.5$) actually becomes asymptotically equal to the pattern of a
single input signal at the center of the cavity $x = 0$.
This coincidence arises from the time-symmetry displayed by these two patters, with the
latter having a recurrence at $\tau/2$ with the same form of the superposition.

\begin{figure*}[t]
 \centering
 \includegraphics[width=\textwidth]{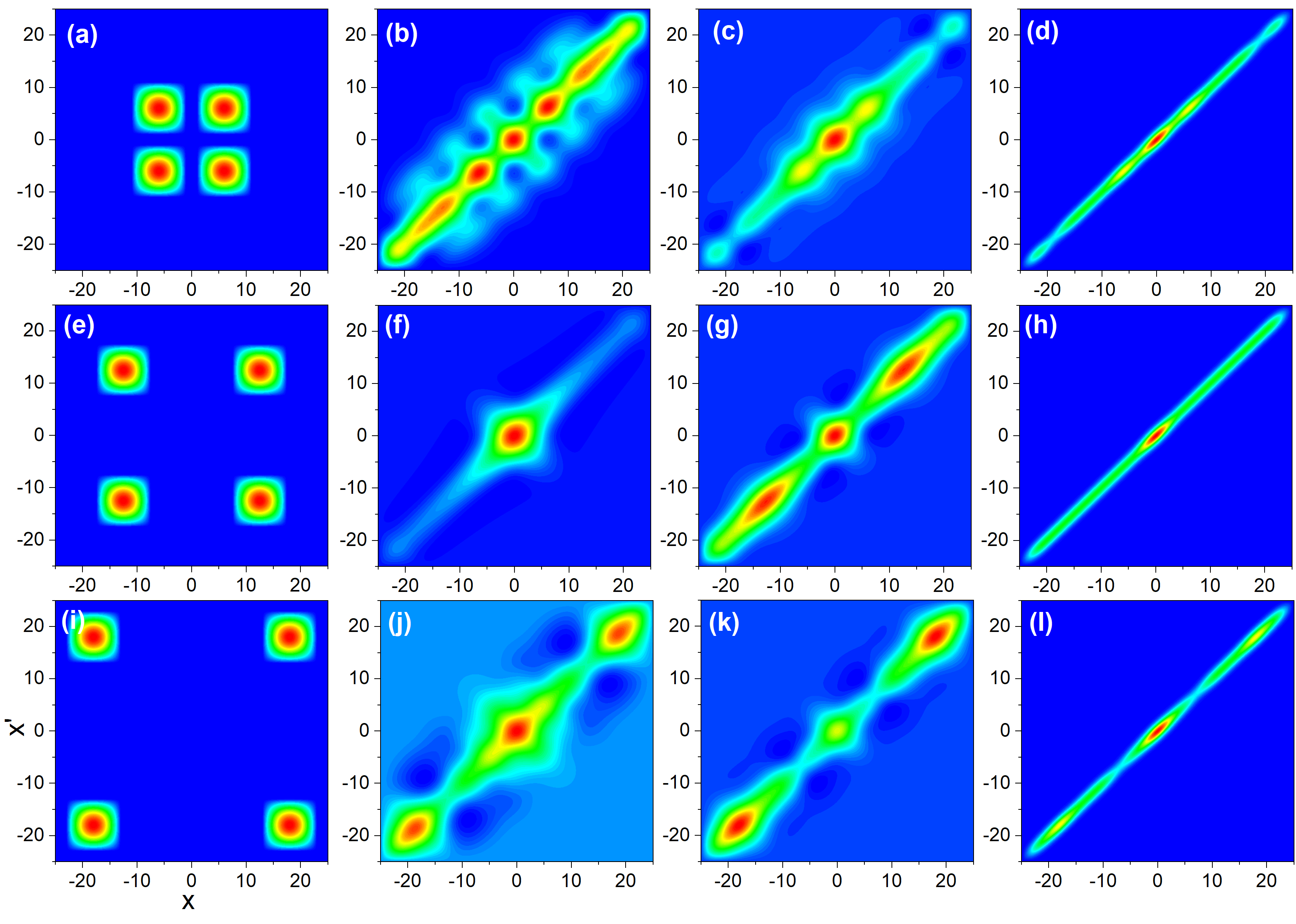}
 \caption{As in Fig.~\ref{fig5}, with $\Lambda$ as given by Eq.~(\ref{localizrate}).}
 \label{fig7}
\end{figure*}

To some extent, one might consider that the above results for $\Lambda=0$ are counter
intuitive.
In principle, one would expect a gradual suppression of all terms outside the main
diagonal of the real part of the density matrix
\cite{miller:JCP-1:2001,brumer-elran:JCP:2004,brumer-elran:JCP:2013,sanz:CJC:2014}.
However, although certainly the rich interference structure outside this diagonal
disappears, certain traits related to the $x-x'$ exchange symmetry still persist.
This is the key aspect that makes necessary in the model here the presence of a
space-dependent damping factor, i.e., a spatial localization rate, $\Lambda$.
This term can be determined analytically \cite{joos:bk:1996} by solving the von
Neumann equation for a free particle acted upon by a scattering-type environment.
Here we have combined this localization rate with the one that is expected from the
cancellation of the interference between different energy modes of the cavity.
The behavior of the real part of the density matrix when this term is added can be
seen in Figs.~\ref{fig6} and \ref{fig7}, which are the respective counterparts of
Figs.~\ref{fig4} and \ref{fig5} for nonvanishing $\Lambda$, with its value given by
(\ref{localizrate}).
As it can be seen, now the model produces the suppression of both energy and space
correlations, thus providing a full phenomenological description to the loss of
coherence inside the cavity.
In the particular case we are dealing with here, though, two-point space correlations do
not directly affect the trajectories (quantum flux), but only the oscillations coming from
interference between different energy modes.
This is consistent with the fact that the localization rate does not appear at all in
the equation of motion (\ref{eom2decoh}).
Now, although this view seems more apparent in the case of the single input signals, in
the case of the superposition it seems that such a term should play a role, as it is
observed in two-slit-type studies where it has also been considered
\cite{sanz:EPJD:2007,luis:AOP:2015}.
If this is not the case, it is precisely because of the spectral decomposition of the
input signal that the equation of motion of the trajectories is based on, which neglects
the fact of two spatially separate signals and reconsiders the initial {\it Ansatz} as a whole,
unlike the two-slit scenario, where no energy decomposition is considered and therefore
the equation of motion is fully based on what happens in the position space.


\subsection{\label{sec33} Decoherence in the energy representation}

\begin{figure}[!t]
 \begin{center}
 \includegraphics[width=\columnwidth]{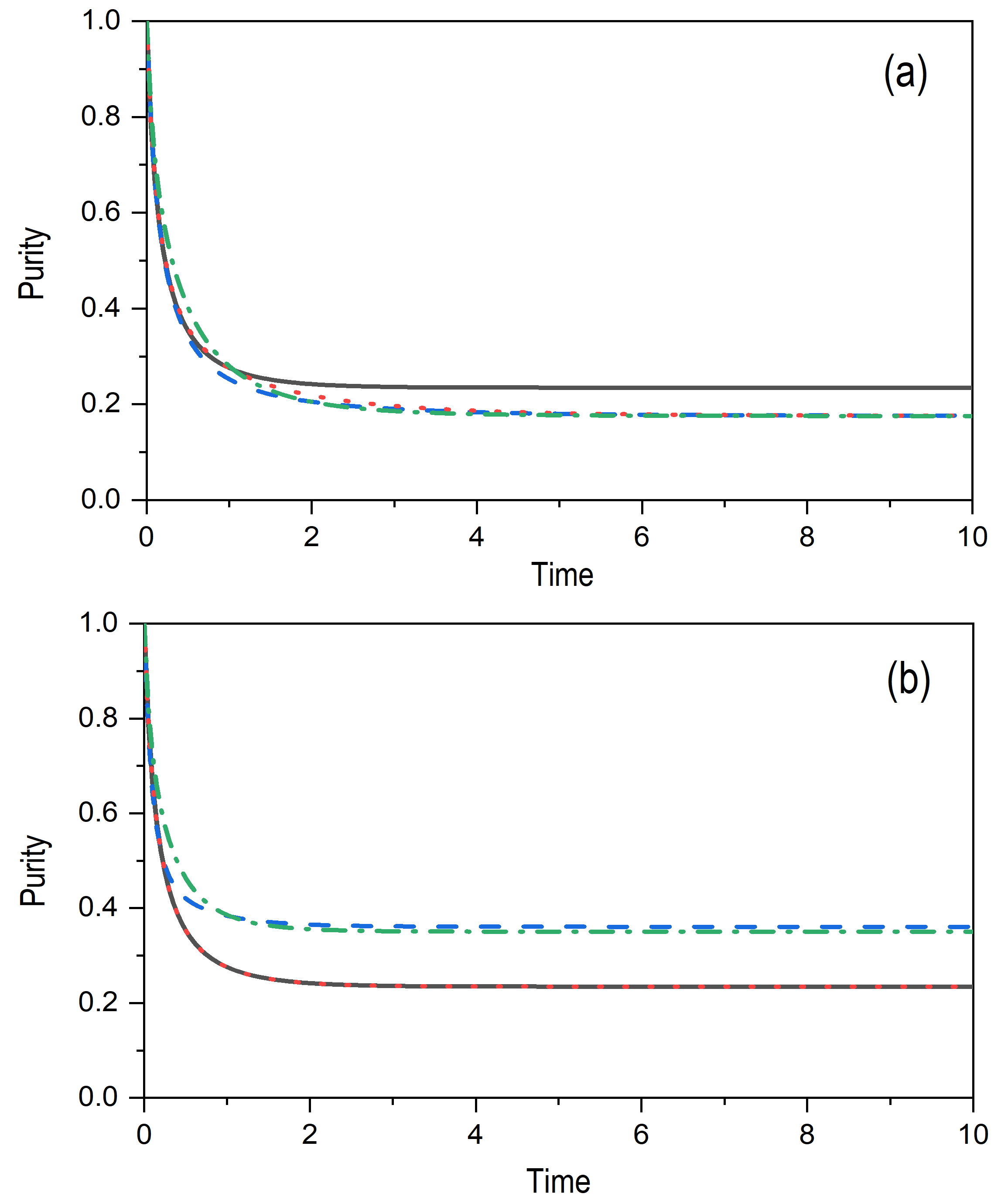}
 \caption{Purity over a time $10\tau$ for (a) single input signals and (b) their symmetric double
  signal counterparts.
  The centers of these inputs are $x_0=6$ (blue dashed line), $x_0=12.5$ (red dotted line) and
  $x_0=18$ (green dash-dotted line) and decrease by one to their corresponding double waves.
  To set a reference to compare with, the result for the input signal centered at $x_0=0$ is
  denoted by a black solid line in both panels.
  In all cases the parameters considered are $\hbar=1$, $m=1$, $L=50$, $w=10$ and
  $\gamma=2/5\pi$.
  All quantities are given in arbitrary units.}
 \label{fig8}
 \end{center}
\end{figure}

To further analyze the implications of the model, now we are going to analyze its
consequences in the energy domain.
To this end, we consider the purity \cite{breuer-bk:2002}, which here acquires the
explicit functional form
\ba
 \chi(t) & = & {\rm Tr} \left( \rho^2 \right)
 \nonumber \\
 & = & \sum_\alpha |c_\alpha|^4 +
 2 \sum_{\alpha' > \alpha} |c_\alpha|^2 |c_{\alpha'}|^2 e^{-2\beta_{\alpha \alpha'}t} ,
 \label{eq:purity}
\ea
with its long-time limit
\be
 \chi_\infty = \sum_\alpha |c_\alpha|^4 .
 \label{eq:purity2}
\ee
This quantity provides us with a reliable measurement of the degree of mixedness
undergone by an initially pure quantum system and, therefore, of the
effects induced by decoherence.
In principle, the density in (\ref{eq:purity}) refers to the reduced
density, i.e., after tracing over the environmental degrees of freedom.
Since the latter is included here in a phenomenological manner, the
density makes direct reference to the density describing the carpet.
Nonetheless, notice that the environmental effects appear explicitly
in the form of the exponential damping factors.

The time-dependence of the purity for an interval equivalent to ten times $\tau$ is
displayed in Fig.~\ref{fig8}(a) for half-cosine input signals centered at different values
of $x_0 \ne 0$.
As it can be noticed, all these cases reach the same asymptotic value of nearly
$\chi_\infty \lesssim 0.2$, below the reference value for $x_0=0$, $\chi_\infty \gtrsim 0.2$.
However, the fall-off slightly differs for each case, which is related to the fact that the
decay factor depends on the frequencies (energy differences) involved as well as the number
of modes and their weight contributing to the corresponding signal.
A similar trend is also observed in the two-signal superpositions, shown in Fig.~\ref{fig8}(b),
with the initial state consisting of the half-cosines centered at the same $x_0$ values and
the mirror image (with respect to $x=0$).
In this case, though, the limiting value is $\chi_\infty \lesssim 0.4$, nearly twice the
value for the single signals, except in the case $x_0 = 12.5$ (red dotted line), which
is the same as for the reference case because of the time-symmetry reasons explained above.

\begin{figure}[!t]
 \centering
 \includegraphics[width=\columnwidth]{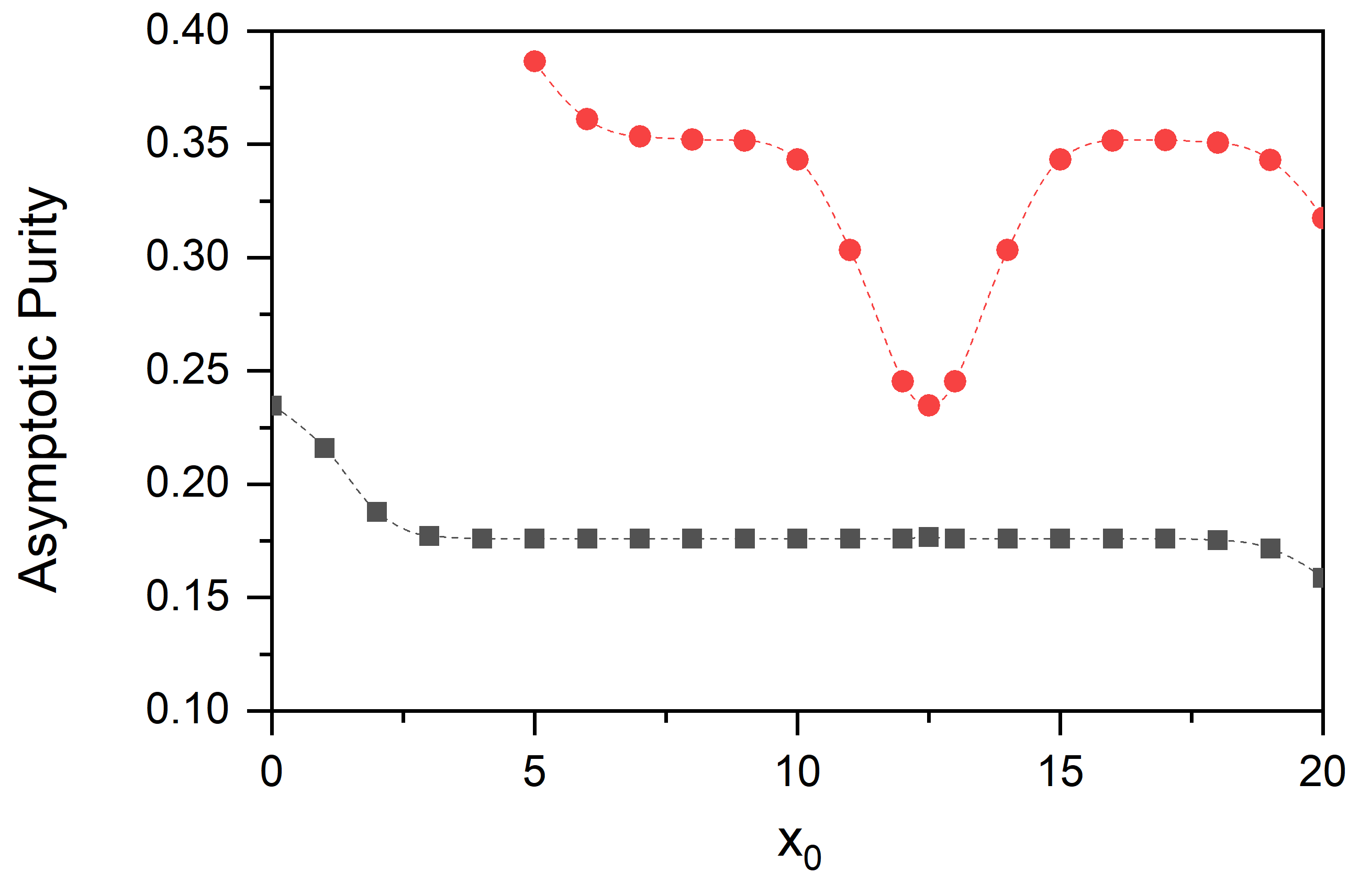}
 \caption{Long-time (asymptotic) values of the purity as a function of $x_0$ for single (black
  squares) and double (red circles) input signals.
  For easier visualization, colored dotted lines joining the data have been added.
  In all cases, the parameters are $m=1$, $\hbar =1$, $L=50$, $w=10$ and $\gamma=2/5\pi$.
  All quantities are given in arbitrary units.}
\label{fig9}
\end{figure}

In order to determine whether the $x_0=0$ case is an exception, we proceed to revise the
asymptotic value (\ref{eq:purity2}) for the whole range between $x_0$ and $20$, which
is the maximum value without truncating the shape of the input signal (note that the cavity
extends to $x=L/2=25$ and the half-width of the input signal is $w=5$).
We proceed the same way with the corresponding superpositions, although this time the
limitation is extended also from below (the minimum $x_0$ is 5) in order to avoid the
overlapping of the two signals.
The results are shown in Fig.~\ref{fig9}, where black squares denote the values considered
for the single half-cosines, while the red circles represent those for the
superpositions.
As can be clearly seen, in both case there is a range where more or less all asymptotic
values remain the same.
Important deviations correspond to the extreme values of $x_0$, where the long-time
purity value undergoes an increase for lower values of $x_0$ and decreases for the
larger ones.
On the other hand, a remarkable feature is the dip observed in the case of the
superpositions, which explains the coincidence between the black solid line and the
red dotted line of Fig.~\ref{fig8}(b): Because the single half-cosine centered at $x_0 = 0$
and the two half-cosine superpositions for $x_0 = L/4 = 12.5$ are equivalent from the
point of view of time symmetry, their purities must coincide, even in the long-time limit.
This is precisely what we observe here if we compare the lowest value of the dip with the
initial value for the single half-cosine (i.e., the one for $x_0 = 0$).
It is thus clear that, in spite of the simplicity of the model, it can provide us with
useful information about the role of symmetries in decoherence processes and hence a
deeper understanding of the carpet-suppression dynamics.

\begin{figure}[!t]
 \centering
 \includegraphics[width=0.9\columnwidth]{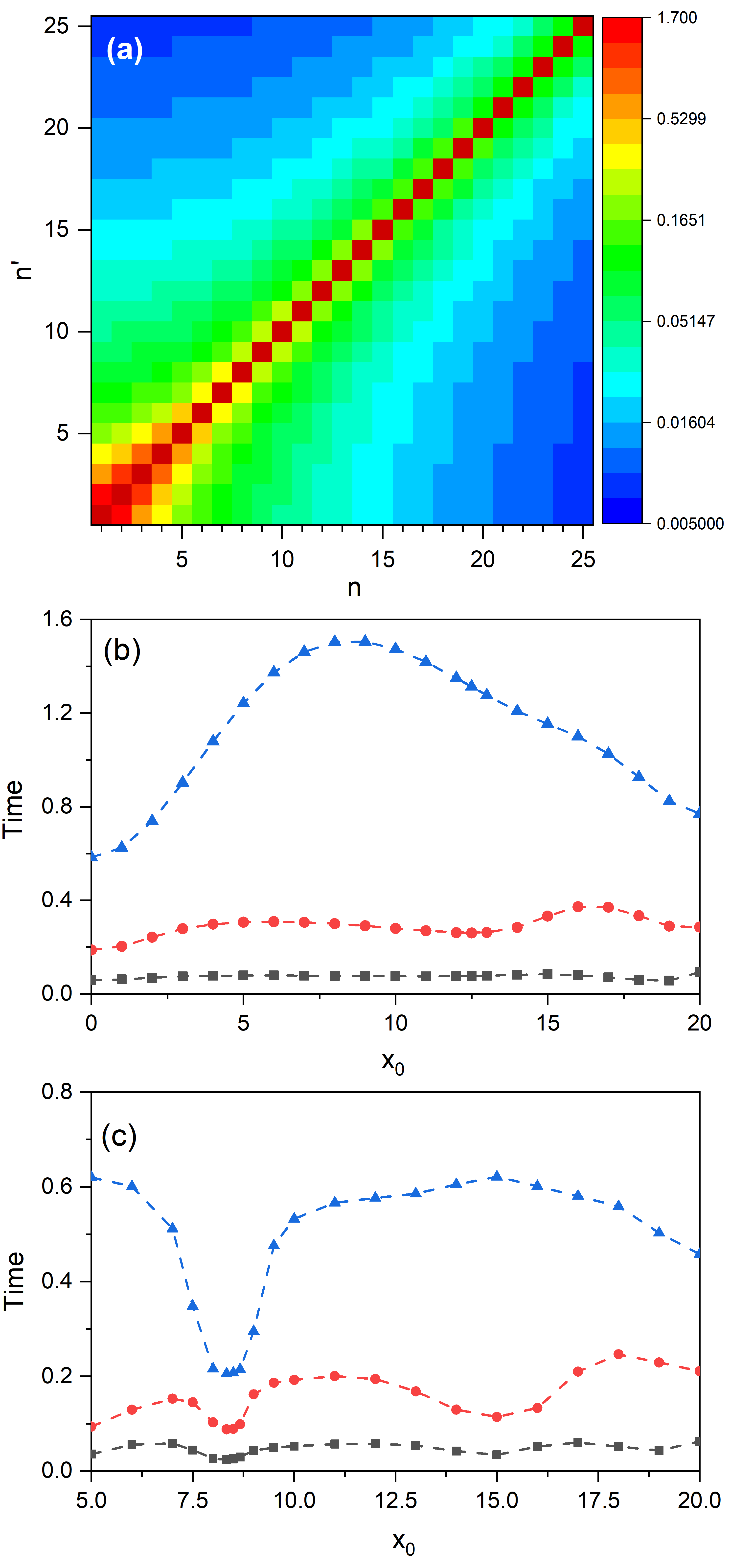}
 \caption{(a) Color map showing the decay times associated with $(\alpha,\alpha')$-pairs of
  modes, where the transition from red to blue denotes shorter and shorter decay times
  (faster and faster decay).
  (b) Characteristic times $t_1$ (black squares), $t_2$ (red circles) and $t_3$ (blue triangles)
  for single input states as a function of the latter center $x_0$.
  (c) Same in (b) but for double input signal.
  For easier visualization, colored dotted lines joining the data have been added.
  In all cases, the parameters are $m=1$, $\hbar =1$, $L=50$, $w=10$ and $\gamma=2/5\pi$.
  All quantities are given in arbitrary units.}
 \label{fig10}
\end{figure}

Some additional information can still be extracted if we analyze the time scales involved
in the decay of the purity.
To that end, it is interesting to consider the decay time-scales related to $(\alpha,\alpha')$
pairs of modes, given by $1/\beta_{\alpha \alpha'}$.
These decay times are independent of the particular input signal selected, since they do not
include the contribution of the particular weight assigned to each pair.
Figure~\ref{fig10}(a) illustrates the distribution of decay times by means of a color map,
where the transition from red to blue denotes shorter and shorter decay times, as expected
for pairs of modes with bigger and bigger energy differences.
Note that in the particular case of the diagonal (dark red color) the decay time becomes infinite
because $\alpha'=\alpha$, this being the main reason why the real part of the density matrix
collapses asymptotically towards this diagonal in the energy representation.
This map thus provides us with a general picture of decay times that does not depend on the
particular choice of the input signal, as mentioned above.

It is clear that the choice of a given input signal is going to have some consequence on
the decay of the carpet, because not only does the time scale $1/\beta_{\alpha \alpha'}$ rule its
decoherence dynamics, but also the particular weight $|c_\alpha| |c_{\alpha'}|$ is going to
play a role depending on whether or not it is important in the superposition (with respect to
other crossed terms or even the populations weights $|c_\alpha|^2$).
To analyze these consequences, now we focus on the shape displayed by the purity and fit it with
a three-time exponentially decaying function
\be
 \chi_{\rm fit} (t) = \chi_0 + \sum_{i=1}^3 \chi_i e^{-(t-t_0)/t_i} .
 \label{eq:ajuste3time}
\ee
The three characteristic times involved here are related to each part of the purity, namely,
the initial falloff ($t_1$), the intermediate turn $t_2$ and slowly decaying tail $t_3$.
Furthermore, all exponentials include a reference onset time $t_0$, the same for all, and the
expression also considers a baseline $\chi_0$, which gives the asymptotic value $\chi_\infty$.
The trend exhibited by the three times in the case of one single signal is displayed in
Fig.~\ref{fig10}(b) in terms of $x_0$.
As it can be noticed, while $t_1$ and $t_2$ are nearly constant in the whole interval, $t_3$
undergoes a remarkable increase as we move towards $x_0 \sim$~8--9 and then it decreases again.
This is because of the larger number of modes involved in the superposition, with nearly
similar weights, which are thus able to support the coherence for longer times.
In the case of two input signals, shown in Fig.~\ref{fig10}(c), the trend is less clear,
although times are much shorter.
Yet it can be noticed that except for a particular value of $x_0$ around 8, where we observe
an important decrease in the three times, although especially in $t_3$, in all other cases
the trend is relatively smooth.
If in the single signal case the maximum for $t_3$ was associated with a rather complex
contribution of modes, the dip here can be justified because the addition of a
mirror-symmetric signal removes many of those contributions, thus generating a rather simple
superposition, much simpler than for other neighboring values of $x_0$.

\begin{figure}[!t]
 \centering
 \includegraphics[width=\columnwidth]{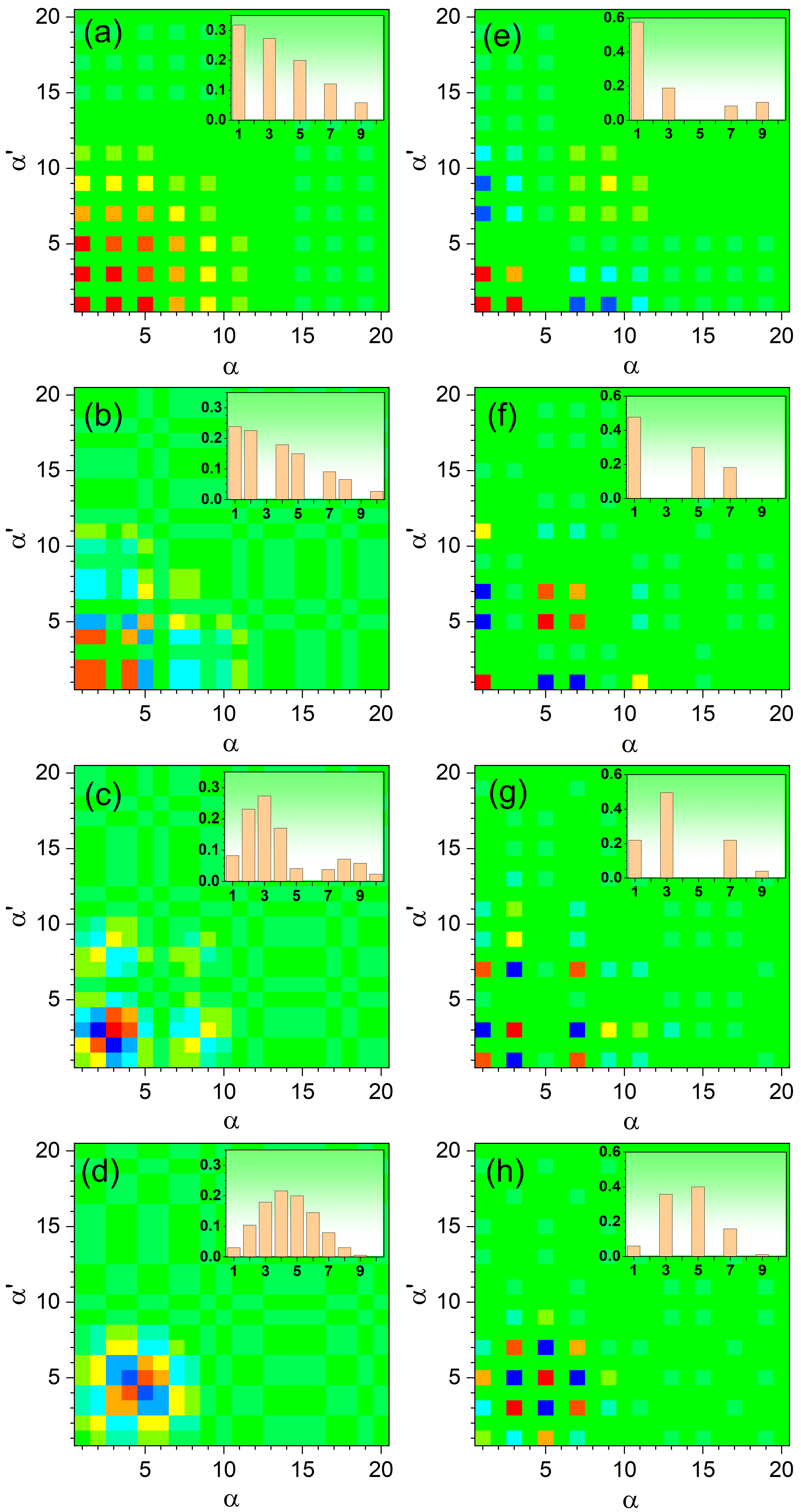}
 \caption{Energy correlation matrix for single (left column) and double
  (right column) input signals.
  For a single input signal (a) $x_0=0$, (b) $x_0=8.3$, (c) $x_0=16.5$, and (d) $x_0=20$.
  For a double input signal (e) $x_0 = 5$, (f) $x_0 = 8.3$, (g) $x_0 = 15$, and (h) $x_0 = 20$.
  The color scale ranges from red denoting higher values (greater than or equal to $0.3$) to blue for lower ones
  (less than or equal to$-0.3$).
  The panel of each inset indicates the population associated with each particular input
  state.
  In all cases, $m=1$, $\hbar=1$, $L=50$, $w=10$ and $\gamma=2/5\pi$.
  All quantities are given in arbitrary units.}
 \label{fig11}
\end{figure}

In order to find an explanation for the different time-scales ruling each decay range of
the purity, let us consider the energy correlation matrix for several single and double
input signals (i.e., in terms of $x_0$), which corresponds to the density matrix at $t=0$,
with elements $\rho_{\alpha \alpha'}(0)= c_\alpha c_{\alpha'}^*$.
These correlations matrices are displayed in Fig.~\ref{fig11} for single input signals
[Fig.~\ref{fig11}(a)--Fig.~\ref{fig11}(d)] and double input signals
[Fig.~\ref{fig11}(e)--Fig.~\ref{fig11}(h)] for different values of $x_0$.
The populations $|c_\alpha|^2$ of each contributing mode are also shown in the inset
of each panel.
Consider first a single input signal
depending on the spatial symmetry displayed by the single input signal as $x_0$ increases,
although in a particular manner.
Up to nearly $x_0 = 8$, the pattern is analogous to the one shown in the inset of Fig.~\ref{fig11}(a),
with a prominent presence of odd modes and residual in the case of the even ones (or
even zero for $x_0=0$), but increasing with $x_0$.
Accordingly, the correlations between neighboring modes increases with $x_0$,
which also leads to an increase of the long-term time $t_3$, as seen in Fig.~\ref{fig10}(a).
Then, between $x_0=8$ and $9$, approximately, we observe paired modes with a lack
or residual presence of correlation between each pair, as can be seen in the inset of Fig.~\ref{fig11}(b).
Because the members of each pair contribute approximately the same, the correlation between
them is going to be rather strong, which leads to a maximum in the long-term time.
Finally, as $x_0$ increases beyond $x_0=9$, more contributions start appearing, increasing the
amount of first, second, third, etc., member correlations, which eventually translates into a
fall in $t_3$.
A similar behavior is also observed in the case of two input signals, although this time,
because of symmetry considerations, there are two positions of particular interest, one
for $x_0 = L/6 \approx 8.3$, where there are no contributions for $n=2$, 3, and 4, thus
implying a strong decay in the correlation matrix, as seen in Fig.~\ref{fig11}(f).
The same happens for $x_0=15$, although this time it is the contribution from the $n=4$, 5
and 6 modes that vanishes, which has an effect on the $t_1$ and $t_2$ times,
as seen in Fig.~\ref{fig10}(b).


\section{\label{sec4} Final remarks}

Although decoherence is commonly regarded as the mechanism by means of which quantum systems
lose their quantumness, thus giving rise to the emergence of a classical world,
the quantum world is never abandoned.
Decoherence is possible because quantum systems interact with other quantum systems and, as
a consequence of such an interaction, when those other surrounding (quantum) systems, known as
the environment, are neglected, the behavior of the system of interest exhibits classical-type
features.
This has given rise in the literature to a series of different phenomenological or effective
decoherence models that try to reproduce the effects of the environment over the system of
interest without explicitly specifying the particular dynamics displayed by each environmental
system.
This is what we generally call the theory of open quantum systems \cite{breuer-bk:2002}, which
is in reality a compendium of effective theories with common grounds that allow us to go from
one to another.

Such theories or models essentially arise from the observation of the behavior displayed by a
quantum system when it is acted upon by an environment.
An immediate effect is the gradual loss of interference features (loss of fringe visibility),
which will be faster or slower depending on the strength of the system-environment interaction.
In this regard, quantum carpets become ideal systems to probe and to evidence the effects of
decoherence \cite{bonifacio,schleich:NJP:2013}, since the intricate carpet-type patterns
exhibited inside a box (a wave guide, regardless of whether the wave is made of light,
electrons, or neutrons) going to be very sensitive to an external action.
In this work we have explored these effects by considering a simple but insightful
coherence damping model based on how more complicated models act on superpositions of energy
states and also on continuous-variable states.
This model describes the action of entangling the system generating the carpets here
with another analogous system, which, when it is traced over, gives rise to the exponential
decay here observed.

What is interesting here is the fact that, because dissipation (thermalization) effects
are not included, a full loss of coherence does not correspond to spatial localization,
as one might expect {\it a priori}.
Rather, it is seen that the annihilation of the coherence terms in the energy density matrix
gives rise to a sort of spreading of energy all along the cavity considered, with some
concentration around the center of the input signal and its mirror-symmetric position with
respect to $x=0$, or around the center in the case of two symmetrically distributed input
signals (initial coherent superpositions of two localized states).
This seemingly unusual long-time limit is actually a product of the bare sum of the densities
associated with each contributing eigenmode.
From this point of view, the oscillatory pattern-type behavior is only a manifestation of how
the coherence terms modulate such a bare backbone.
When this effect is visualized in terms of Bohmian trajectories, a transition from a highly
oscillatory motion to a rather smooth behavior is observed, with a long-time limit being described
by motionless trajectories.
This is a nice manifestation of the important connection between Bohmian mechanics and an underlying
locally varying phase field: As soon as the phase field does not change any more, all Bohmian motion
disappears.
This situation resembles the widely known stationarity associated with non-degenerate eigenvalues,
which has been debated for a long time in the literature \cite{holland-bk}.
It is worth mentioning that at present these behaviors, fully based on Markovian
considerations, are being investigated also in terms of a more robust two-party
entanglement model.
It is expected that the model will render in a natural fashion the damping factor in positions, but
also a counterpart in the energy representation, thus offering an alternative perspective on the
coherence transference between parties and its role in decoherence processes.

Another interesting feature that has been observed is that not only the coherence
among different energy eigenstates must disappear in order to faithfully reproduce the effects
of decoherence.
If instead of the probability density we look at the real part of the density matrix (the
imaginary part is valid as well), we notice that important structures out of the main diagonal
still persist.
The reason is that the removal of these contributions requires an additional element related
to the loss of point-to-point space correlations, which cannot be readily seen in more
refined models.
When we introduce a damping term in this direction, all contributions out of the main diagonal
gradually disappear.
Note that this is a feature that cannot be observed by simply inspecting the probability
density, because although interferential traits are seen to disappear with time, the same
is not seen at all in the density matrix in space, since all those correlations disappear
when we set $x=x'$.

Furthermore, in order to determine in a more quantitative manner the effectiveness of decoherence
in terms of the initial input state considered, analysis of the purity and the density matrix
in the energy representation has also been carried out.
In this regard it is worth mentioning that, when the input states have an even parity, the
loss of coherence takes place more rapidly than in the case of asymmetric states due to the
many more cavity modes that such states can be projected on. 
To quantify the effect, we analyzed the time scales involved in each case,
finding that there are three decay time-scales ruling the behavior of the purity in the
short, medium, and long terms, respectively.

Finally, it is worth emphasizing that the analysis presented here, which includes
different supplementary tools (densities in space and momentum, density matrices and Bohmian
trajectories), can be equally applied to confinement of matter waves and to light pulses pumped
into resonant cavities or multimode interference devices, where the generation of twin pulses
is based on the phenomenon of recurrences.
In these cases, taking advantage of the isomorphism between the Schr\"odinger equation and
the paraxial Helmholtz equation \cite{sanz:JOSAA:2012}, the full analysis presented here for
a matter wave could be profitably switched to the analysis of confined light.


\begin{acknowledgments}

Financial support from the Agencia Estatal de Investigaci\'on (AEI) and the
European Regional Development Fund (ERDF) through Grant No.\ FIS2016-76110-P
is acknowledged.

\end{acknowledgments}



\begin{thebibliography}{69}%
\makeatletter
\providecommand \@ifxundefined [1]{%
 \@ifx{#1\undefined}
}%
\providecommand \@ifnum [1]{%
 \ifnum #1\expandafter \@firstoftwo
 \else \expandafter \@secondoftwo
 \fi
}%
\providecommand \@ifx [1]{%
 \ifx #1\expandafter \@firstoftwo
 \else \expandafter \@secondoftwo
 \fi
}%
\providecommand \natexlab [1]{#1}%
\providecommand \enquote  [1]{``#1''}%
\providecommand \bibnamefont  [1]{#1}%
\providecommand \bibfnamefont [1]{#1}%
\providecommand \citenamefont [1]{#1}%
\providecommand \href@noop [0]{\@secondoftwo}%
\providecommand \href [0]{\begingroup \@sanitize@url \@href}%
\providecommand \@href[1]{\@@startlink{#1}\@@href}%
\providecommand \@@href[1]{\endgroup#1\@@endlink}%
\providecommand \@sanitize@url [0]{\catcode `\\12\catcode `\$12\catcode
  `\&12\catcode `\#12\catcode `\^12\catcode `\_12\catcode `\%12\relax}%
\providecommand \@@startlink[1]{}%
\providecommand \@@endlink[0]{}%
\providecommand \url  [0]{\begingroup\@sanitize@url \@url }%
\providecommand \@url [1]{\endgroup\@href {#1}{\urlprefix }}%
\providecommand \urlprefix  [0]{URL }%
\providecommand \Eprint [0]{\href }%
\providecommand \doibase [0]{https://doi.org/}%
\providecommand \selectlanguage [0]{\@gobble}%
\providecommand \bibinfo  [0]{\@secondoftwo}%
\providecommand \bibfield  [0]{\@secondoftwo}%
\providecommand \translation [1]{[#1]}%
\providecommand \BibitemOpen [0]{}%
\providecommand \bibitemStop [0]{}%
\providecommand \bibitemNoStop [0]{.\EOS\space}%
\providecommand \EOS [0]{\spacefactor3000\relax}%
\providecommand \BibitemShut  [1]{\csname bibitem#1\endcsname}%
\let\auto@bib@innerbib\@empty
\bibitem [{\citenamefont {Breuer}\ and\ \citenamefont
  {Petruccione}(2002)}]{breuer-bk:2002}%
  \BibitemOpen
  \bibfield  {author} {\bibinfo {author} {\bibfnamefont {H.-P.}\ \bibnamefont
  {Breuer}}\ and\ \bibinfo {author} {\bibfnamefont {F.}~\bibnamefont
  {Petruccione}},\ }\href@noop {} {\emph {\bibinfo {title} {The Theory of Open
  Quantum Systems}}}\ (\bibinfo  {publisher} {Oxford University Press},\
  \bibinfo {address} {Oxford},\ \bibinfo {year} {2002})\BibitemShut {NoStop}%
\bibitem [{\citenamefont {Joos}\ and\ \citenamefont
  {Zeh}(1985)}]{joos-zeh:ZPhysB:1985}%
  \BibitemOpen
  \bibfield  {author} {\bibinfo {author} {\bibfnamefont {E.}~\bibnamefont
  {Joos}}\ and\ \bibinfo {author} {\bibfnamefont {H.~D.}\ \bibnamefont {Zeh}},\
  }\bibfield  {title} {\bibinfo {title} {The emergence of classical properties
  through interaction with the environment},\ }\href@noop {} {\bibfield
  {journal} {\bibinfo  {journal} {Z. Phys. B}\ }\textbf {\bibinfo {volume}
  {59}},\ \bibinfo {pages} {223} (\bibinfo {year} {1985})}\BibitemShut
  {NoStop}%
\bibitem [{\citenamefont {Zurek}(1991)}]{zurek:PhysToday:1991}%
  \BibitemOpen
  \bibfield  {author} {\bibinfo {author} {\bibfnamefont {W.~H.}\ \bibnamefont
  {Zurek}},\ }\bibfield  {title} {\bibinfo {title} {Decoherence and the
  transition from quantum to classical},\ }\href@noop {} {\bibfield  {journal}
  {\bibinfo  {journal} {Phys. Today}\ }\textbf {\bibinfo {volume} {44(10)}},\
  \bibinfo {pages} {36} (\bibinfo {year} {1991})}\BibitemShut {NoStop}%
\bibitem [{\citenamefont {Zurek}(2002)}]{zurek:PhysToday-rev:2003}%
  \BibitemOpen
  \bibfield  {author} {\bibinfo {author} {\bibfnamefont {W.~H.}\ \bibnamefont
  {Zurek}},\ }\bibfield  {title} {\bibinfo {title} {Decoherence and the
  transition from quantum to classical---Revisited},\ }\href@noop {}
  {\bibfield  {journal} {\bibinfo  {journal} {Los Alamos Science}\ }\textbf
  {\bibinfo {volume} {27}},\ \bibinfo {pages} {2} (\bibinfo {year}
  {2002})}\BibitemShut {NoStop}%
\bibitem [{\citenamefont {Blanchard}\ and\ \citenamefont
  {Olkiewicz}(2003)}]{blanchardolkiewicz}%
  \BibitemOpen
  \bibfield  {author} {\bibinfo {author} {\bibfnamefont {P.}~\bibnamefont
  {Blanchard}}\ and\ \bibinfo {author} {\bibfnamefont {R.}~\bibnamefont
  {Olkiewicz}},\ }\bibfield  {title} {\bibinfo {title} {Decoherence induced
  transition from quantum to classical dynamics},\ }\href@noop {} {\bibfield
  {journal} {\bibinfo  {journal} {Rev. Math. Phys.}\ }\textbf {\bibinfo
  {volume} {15}},\ \bibinfo {pages} {217} (\bibinfo {year} {2003})}\BibitemShut
  {NoStop}%
\bibitem [{\citenamefont {Giulini}\ \emph {et~al.}(rlin)\citenamefont
  {Giulini}, \citenamefont {Joos}, \citenamefont {Kiefer}, \citenamefont
  {Kupsch}, \citenamefont {Stamatescu},\ and\ \citenamefont
  {Zeh}}]{giulini-bk}%
  \BibitemOpen
  \bibfield  {author} {\bibinfo {author} {\bibfnamefont {D.}~\bibnamefont
  {Giulini}}, \bibinfo {author} {\bibfnamefont {E.}~\bibnamefont {Joos}},
  \bibinfo {author} {\bibfnamefont {C.}~\bibnamefont {Kiefer}}, \bibinfo
  {author} {\bibfnamefont {J.}~\bibnamefont {Kupsch}}, \bibinfo {author}
  {\bibfnamefont {I.-O.}\ \bibnamefont {Stamatescu}},\ and\ \bibinfo {author}
  {\bibfnamefont {H.~D.}\ \bibnamefont {Zeh}},\ }\href@noop {} {\emph {\bibinfo
  {title} {Decoherence and the Appearance of a Classical World in Quantum
  Theory}}},\ \bibinfo {edition} {2nd}\ ed.\ (\bibinfo  {publisher}
  {Springer},\ \bibinfo {address} {1996},\ \bibinfo {year}
  {Berlin})\BibitemShut {NoStop}%
\bibitem [{\citenamefont {Schlosshauer}(2004)}]{schlosshauer:RMP:2004}%
  \BibitemOpen
  \bibfield  {author} {\bibinfo {author} {\bibfnamefont {M.}~\bibnamefont
  {Schlosshauer}},\ }\bibfield  {title} {\bibinfo {title} {Decoherence, the
  measurement problem, and interpretations of quantum mechanics},\ }\href@noop
  {} {\bibfield  {journal} {\bibinfo  {journal} {Rev. Mod. Phys.}\ }\textbf
  {\bibinfo {volume} {76}},\ \bibinfo {pages} {1267} (\bibinfo {year}
  {2004})}\BibitemShut {NoStop}%
\bibitem [{\citenamefont {Schlosshauer}(2007)}]{schlosshauer-bk:2007}%
  \BibitemOpen
  \bibfield  {author} {\bibinfo {author} {\bibfnamefont {M.}~\bibnamefont
  {Schlosshauer}},\ }\href@noop {} {\emph {\bibinfo {title} {Decoherence and
  the Quantum-to-Classical Transition}}}\ (\bibinfo  {publisher} {Springer},\
  \bibinfo {address} {Berlin},\ \bibinfo {year} {2007})\BibitemShut {NoStop}%
\bibitem [{\citenamefont {Schlosshauer}(2019)}]{schlosshauerdecoh}%
  \BibitemOpen
  \bibfield  {author} {\bibinfo {author} {\bibfnamefont {M.}~\bibnamefont
  {Schlosshauer}},\ }\bibfield  {title} {\bibinfo {title} {Quantum
  decoherence},\ }\href@noop {} {\bibfield  {journal} {\bibinfo  {journal}
  {Phys. Rep.}\ }\textbf {\bibinfo {volume} {831}},\ \bibinfo {pages} {1}
  (\bibinfo {year} {2019})}\BibitemShut {NoStop}%
\bibitem [{\citenamefont {Khaetskii}\ \emph {et~al.}(2002)\citenamefont
  {Khaetskii}, \citenamefont {Loss},\ and\ \citenamefont
  {Glazman}}]{khaetskii:PRL:2002}%
  \BibitemOpen
  \bibfield  {author} {\bibinfo {author} {\bibfnamefont {A.~V.}\ \bibnamefont
  {Khaetskii}}, \bibinfo {author} {\bibfnamefont {D.}~\bibnamefont {Loss}},\
  and\ \bibinfo {author} {\bibfnamefont {L.}~\bibnamefont {Glazman}},\
  }\bibfield  {title} {\bibinfo {title} {Electron spin decoherence in quantum
  dots due to interaction with nuclei},\ }\href@noop {} {\bibfield  {journal}
  {\bibinfo  {journal} {Phys. Rev. Lett.}\ }\textbf {\bibinfo {volume} {88}},\
  \bibinfo {pages} {186802} (\bibinfo {year} {2002})}\BibitemShut {NoStop}%
\bibitem [{\citenamefont {Mathew}\ and\ \citenamefont
  {Nandy}(2013)}]{mathew:MPLB:2013}%
  \BibitemOpen
  \bibfield  {author} {\bibinfo {author} {\bibfnamefont {A.}~\bibnamefont
  {Mathew}}\ and\ \bibinfo {author} {\bibfnamefont {M.~K.}\ \bibnamefont
  {Nandy}},\ }\bibfield  {title} {\bibinfo {title} {Decoherence study of
  electron spin states in quantum dots using a simplistic model},\ }\href@noop
  {} {\bibfield  {journal} {\bibinfo  {journal} {Mod. Phys. Lett. B}\ }\textbf
  {\bibinfo {volume} {27}},\ \bibinfo {pages} {1350119} (\bibinfo {year}
  {2013})}\BibitemShut {NoStop}%
\bibitem [{\citenamefont {Altaisky}\ \emph {et~al.}(2016)\citenamefont
  {Altaisky}, \citenamefont {Zolnikova}, \citenamefont {Kaputkina},
  \citenamefont {Krylov}, \citenamefont {Lozovik},\ and\ \citenamefont
  {Dattani}}]{altaisky:EPJWebConf:2016}%
  \BibitemOpen
  \bibfield  {author} {\bibinfo {author} {\bibfnamefont {M.~V.}\ \bibnamefont
  {Altaisky}}, \bibinfo {author} {\bibfnamefont {N.~N.}\ \bibnamefont
  {Zolnikova}}, \bibinfo {author} {\bibfnamefont {N.~E.}\ \bibnamefont
  {Kaputkina}}, \bibinfo {author} {\bibfnamefont {V.~A.}\ \bibnamefont
  {Krylov}}, \bibinfo {author} {\bibfnamefont {Y.~E.}\ \bibnamefont
  {Lozovik}},\ and\ \bibinfo {author} {\bibfnamefont {N.~S.}\ \bibnamefont
  {Dattani}},\ }\bibfield  {title} {\bibinfo {title} {Decoherence and
  entanglement simulation in a model of quantum neural wetwork based on quantum
  dots},\ }\href@noop {} {\bibfield  {journal} {\bibinfo  {journal} {EPJ Web of
  Conferences}\ }\textbf {\bibinfo {volume} {108}},\ \bibinfo {pages} {02006}
  (\bibinfo {year} {2016})}\BibitemShut {NoStop}%
\bibitem [{\citenamefont {Flitney}\ and\ \citenamefont
  {Abbott}(2005)}]{flitneyabbot}%
  \BibitemOpen
  \bibfield  {author} {\bibinfo {author} {\bibfnamefont {A.~P.}\ \bibnamefont
  {Flitney}}\ and\ \bibinfo {author} {\bibfnamefont {D.}~\bibnamefont
  {Abbott}},\ }\bibfield  {title} {\bibinfo {title} {Quantum games with
  decoherence},\ }\href@noop {} {\bibfield  {journal} {\bibinfo  {journal} {J.
  Phys. A: Math. Gen.}\ }\textbf {\bibinfo {volume} {38}},\ \bibinfo {pages}
  {449} (\bibinfo {year} {2005})}\BibitemShut {NoStop}%
\bibitem [{\citenamefont {Chen}\ \emph {et~al.}(2003)\citenamefont {Chen},
  \citenamefont {Ang}, \citenamefont {Kiang}, \citenamefont {Kwek},\ and\
  \citenamefont {Lo}}]{prisoner}%
  \BibitemOpen
  \bibfield  {author} {\bibinfo {author} {\bibfnamefont {L.~K.}\ \bibnamefont
  {Chen}}, \bibinfo {author} {\bibfnamefont {H.~L.}\ \bibnamefont {Ang}},
  \bibinfo {author} {\bibfnamefont {D.}~\bibnamefont {Kiang}}, \bibinfo
  {author} {\bibfnamefont {L.~C.}\ \bibnamefont {Kwek}},\ and\ \bibinfo
  {author} {\bibfnamefont {C.~F.}\ \bibnamefont {Lo}},\ }\bibfield  {title}
  {\bibinfo {title} {Quantum prisoner dilemma under decoherence},\ }\href@noop
  {} {\bibfield  {journal} {\bibinfo  {journal} {Phys. Lett. A}\ }\textbf
  {\bibinfo {volume} {316}},\ \bibinfo {pages} {317} (\bibinfo {year}
  {2003})}\BibitemShut {NoStop}%
\bibitem [{\citenamefont {Kendon}\ and\ \citenamefont
  {Tregenna}(2003)}]{kendon:PRA:2003}%
  \BibitemOpen
  \bibfield  {author} {\bibinfo {author} {\bibfnamefont {V.}~\bibnamefont
  {Kendon}}\ and\ \bibinfo {author} {\bibfnamefont {B.}~\bibnamefont
  {Tregenna}},\ }\bibfield  {title} {\bibinfo {title} {Decoherence can be
  useful in quantum walks},\ }\href@noop {} {\bibfield  {journal} {\bibinfo
  {journal} {Phys. Rev. A}\ }\textbf {\bibinfo {volume} {67}},\ \bibinfo
  {pages} {042315} (\bibinfo {year} {2003})}\BibitemShut {NoStop}%
\bibitem [{\citenamefont {Yin}\ \emph {et~al.}(2008)\citenamefont {Yin},
  \citenamefont {Katsanos},\ and\ \citenamefont {Evangelou}}]{yin:PRA:2008}%
  \BibitemOpen
  \bibfield  {author} {\bibinfo {author} {\bibfnamefont {Y.}~\bibnamefont
  {Yin}}, \bibinfo {author} {\bibfnamefont {D.~E.}\ \bibnamefont {Katsanos}},\
  and\ \bibinfo {author} {\bibfnamefont {S.~N.}\ \bibnamefont {Evangelou}},\
  }\bibfield  {title} {\bibinfo {title} {Quantum walks on a random
  environment},\ }\href@noop {} {\bibfield  {journal} {\bibinfo  {journal}
  {Phys. Rev. A}\ }\textbf {\bibinfo {volume} {77}},\ \bibinfo {pages}
  {022302} (\bibinfo {year} {2008})}\BibitemShut {NoStop}%
\bibitem [{\citenamefont {Milburn}(1991)}]{milburn1}%
  \BibitemOpen
  \bibfield  {author} {\bibinfo {author} {\bibfnamefont {G.~J.}\ \bibnamefont
  {Milburn}},\ }\bibfield  {title} {\bibinfo {title} {Intrinsic decoherence in
  quantum mechanics},\ }\href@noop {} {\bibfield  {journal} {\bibinfo
  {journal} {Phys. Rev. A}\ }\textbf {\bibinfo {volume} {44}},\ \bibinfo
  {pages} {5401} (\bibinfo {year} {1991})}\BibitemShut {NoStop}%
\bibitem [{\citenamefont {Leman}(1998)}]{leman:CTP:1998}%
  \BibitemOpen
  \bibfield  {author} {\bibinfo {author} {\bibfnamefont {K.}~\bibnamefont
  {Leman}},\ }\bibfield  {title} {\bibinfo {title} {Decoherence induced by
  discontinuous and stochastic unitary evolution of qubits in quantum
  computers},\ }\href@noop {} {\bibfield  {journal} {\bibinfo  {journal}
  {Commun. Theor. Phys.}\ }\textbf {\bibinfo {volume} {29}},\ \bibinfo {pages}
  {169} (\bibinfo {year} {1998})}\BibitemShut {NoStop}%
\bibitem [{\citenamefont {Kimm}\ and\ \citenamefont
  {Kwon}(2002)}]{kimm:PRA:2002}%
  \BibitemOpen
  \bibfield  {author} {\bibinfo {author} {\bibfnamefont {K.}~\bibnamefont
  {Kimm}}\ and\ \bibinfo {author} {\bibfnamefont {H.~H.}\ \bibnamefont
  {Kwon}},\ }\bibfield  {title} {\bibinfo {title} {Decoherence of the quantum
  gate in Milburn's model of decoherence},\ }\href@noop {} {\bibfield
  {journal} {\bibinfo  {journal} {Phys. Rev. A}\ }\textbf {\bibinfo {volume}
  {65}},\ \bibinfo {pages}
  {022311} (\bibinfo {year} {2002})}\BibitemShut {NoStop}%
\bibitem [{\citenamefont {Wu}\ \emph {et~al.}(2017)\citenamefont {Wu},
  \citenamefont {Deng}, \citenamefont {Li},\ and\ \citenamefont
  {Sarma}}]{wudenglidassarma}%
  \BibitemOpen
  \bibfield  {author} {\bibinfo {author} {\bibfnamefont {Y.~L.}\ \bibnamefont
  {Wu}}, \bibinfo {author} {\bibfnamefont {D.~L.}\ \bibnamefont {Deng}},
  \bibinfo {author} {\bibfnamefont {X.~P.}\ \bibnamefont {Li}},\ and\ \bibinfo
  {author} {\bibfnamefont {S.~D.}\ \bibnamefont {Sarma}},\ }\bibfield  {title}
  {\bibinfo {title} {Intrinsic decoherence in isolated quantum systems},\
  }\href@noop {} {\bibfield  {journal} {\bibinfo  {journal} {Phys. Rev. B}\
  }\textbf {\bibinfo {volume} {95}},\ \bibinfo {pages}
  {014202} (\bibinfo {year} {2017})}\BibitemShut
  {NoStop}%
\bibitem [{\citenamefont {Ziman}\ and\ \citenamefont
  {Bu\v{z}ek}(2005)}]{ziman:PRA:2005}%
  \BibitemOpen
  \bibfield  {author} {\bibinfo {author} {\bibfnamefont {M.}~\bibnamefont
  {Ziman}}\ and\ \bibinfo {author} {\bibfnamefont {V.}~\bibnamefont
  {Bu\v{z}ek}},\ }\bibfield  {title} {\bibinfo {title} {All (qubit)
  decoherences: Complete characterization and physical implementation},\
  }\href@noop {} {\bibfield  {journal} {\bibinfo  {journal} {Phys. Rev. A}\
  }\textbf {\bibinfo {volume} {72}},\ \bibinfo {pages} {022110} (\bibinfo
  {year} {2005})}\BibitemShut {NoStop}%
\bibitem [{\citenamefont {Czerwinski}(2016)}]{czerwinski:IJTP:2016}%
  \BibitemOpen
  \bibfield  {author} {\bibinfo {author} {\bibfnamefont {A.}~\bibnamefont
  {Czerwinski}},\ }\bibfield  {title} {\bibinfo {title} {Applications of the
  stroboscopic tomography to selected 2-level decoherence models},\ }\href@noop
  {} {\bibfield  {journal} {\bibinfo  {journal} {Int. J. Theor. Phys.}\
  }\textbf {\bibinfo {volume} {55}},\ \bibinfo {pages} {658} (\bibinfo {year}
  {2016})}\BibitemShut {NoStop}%
\bibitem [{\citenamefont {Floss}\ \emph {et~al.}(2019)\citenamefont {Floss},
  \citenamefont {Lemell}, \citenamefont {Yabana},\ and\ \citenamefont
  {Burgd\"orfer}}]{floss:PRB:2019}%
  \BibitemOpen
  \bibfield  {author} {\bibinfo {author} {\bibfnamefont {I.}~\bibnamefont
  {Floss}}, \bibinfo {author} {\bibfnamefont {C.}~\bibnamefont {Lemell}},
  \bibinfo {author} {\bibfnamefont {K.}~\bibnamefont {Yabana}},\ and\ \bibinfo
  {author} {\bibfnamefont {J.}~\bibnamefont {Burgd\"orfer}},\ }\bibfield
  {title} {\bibinfo {title} {Incorporating decoherence into solid-state
  time-dependent density functional theory},\ }\href@noop {} {\bibfield
  {journal} {\bibinfo  {journal} {Phys. Rev. B}\ }\textbf {\bibinfo {volume}
  {99}},\ \bibinfo {pages} {224301} (\bibinfo {year} {2019})}\BibitemShut
  {NoStop}%
\bibitem [{\citenamefont {Longhi}(2013)}]{longhi}%
  \BibitemOpen
  \bibfield  {author} {\bibinfo {author} {\bibfnamefont {S.}~\bibnamefont
  {Longhi}},\ }\bibfield  {title} {\bibinfo {title} {Quantum simulation of
  decoherence in optical waveguide lattices},\ }\href@noop {} {\bibfield
  {journal} {\bibinfo  {journal} {Opt. Lett.}\ }\textbf {\bibinfo {volume}
  {38}},\ \bibinfo {pages} {4884} (\bibinfo {year} {2013})}\BibitemShut
  {NoStop}%
\bibitem [{\citenamefont {Chen}\ and\ \citenamefont
  {Gao}(2013)}]{chen:CTP:2013}%
  \BibitemOpen
  \bibfield  {author} {\bibinfo {author} {\bibfnamefont {C.}~\bibnamefont
  {Chen}}\ and\ \bibinfo {author} {\bibfnamefont {Y.~B.}\ \bibnamefont {Gao}},\
  }\bibfield  {title} {\bibinfo {title} {Quantum decoherence of charge qubit
  coupled to nonlinear nanomechanical resonator},\ }\href@noop {} {\bibfield
  {journal} {\bibinfo  {journal} {Commun. Theor. Phys.}\ }\textbf {\bibinfo
  {volume} {60}},\ \bibinfo {pages} {531} (\bibinfo {year} {2013})}\BibitemShut
  {NoStop}%
\bibitem [{\citenamefont {Dlamini}\ \emph {et~al.}(2018)\citenamefont
  {Dlamini}, \citenamefont {Francis}, \citenamefont {Zhang}, \citenamefont
  {\c{S}. K.~\"Ozdemir}, \citenamefont {Chormaic}, \citenamefont
  {Petruccione},\ and\ \citenamefont {Tame}}]{petruccione:PRAppl:2018}%
  \BibitemOpen
  \bibfield  {author} {\bibinfo {author} {\bibfnamefont {S.~G.}\ \bibnamefont
  {Dlamini}}, \bibinfo {author} {\bibfnamefont {J.~T.}\ \bibnamefont
  {Francis}}, \bibinfo {author} {\bibfnamefont {X.}~\bibnamefont {Zhang}},
  \bibinfo {author} {\bibnamefont {\c{S}. K.~\"Ozdemir}}, \bibinfo {author}
  {\bibfnamefont {S.~N.}\ \bibnamefont {Chormaic}}, \bibinfo {author}
  {\bibfnamefont {F.}~\bibnamefont {Petruccione}},\ and\ \bibinfo {author}
  {\bibfnamefont {M.~S.}\ \bibnamefont {Tame}},\ }\bibfield  {title} {\bibinfo
  {title} {Probing decoherence in plasmonic waveguides in the quantum regime},\
  }\href@noop {} {\bibfield  {journal} {\bibinfo  {journal} {Phys. Rev. Appl.}\
  }\textbf {\bibinfo {volume} {9}},\ \bibinfo {pages} {024003} (\bibinfo
  {year} {2018})}\BibitemShut {NoStop}%
\bibitem [{\citenamefont {Schroll}\ \emph {et~al.}(2003)\citenamefont
  {Schroll}, \citenamefont {Belzig},\ and\ \citenamefont
  {Bruder}}]{schroll:PRA:2003}%
  \BibitemOpen
  \bibfield  {author} {\bibinfo {author} {\bibfnamefont {C.}~\bibnamefont
  {Schroll}}, \bibinfo {author} {\bibfnamefont {W.}~\bibnamefont {Belzig}},\
  and\ \bibinfo {author} {\bibfnamefont {C.}~\bibnamefont {Bruder}},\
  }\bibfield  {title} {\bibinfo {title} {Decoherence of cold atomic gases in
  magnetic microtraps},\ }\href@noop {} {\bibfield  {journal} {\bibinfo
  {journal} {Phys. Rev. A}\ }\textbf {\bibinfo {volume} {68}},\ \bibinfo
  {pages} {043618} (\bibinfo {year} {2003})}\BibitemShut {NoStop}%
\bibitem [{\citenamefont {Dehdashti}\ \emph {et~al.}(2013)\citenamefont
  {Dehdashti}, \citenamefont {Mahdifar}, \citenamefont {Harouni},\ and\
  \citenamefont {Roknizadeh}}]{dehdashti}%
  \BibitemOpen
  \bibfield  {author} {\bibinfo {author} {\bibfnamefont {S.}~\bibnamefont
  {Dehdashti}}, \bibinfo {author} {\bibfnamefont {A.}~\bibnamefont {Mahdifar}},
  \bibinfo {author} {\bibfnamefont {M.~B.}\ \bibnamefont {Harouni}},\ and\
  \bibinfo {author} {\bibfnamefont {R.}~\bibnamefont {Roknizadeh}},\ }\bibfield
   {title} {\bibinfo {title} {Decoherence of spin-deformed bosonic model},\
  }\href@noop {} {\bibfield  {journal} {\bibinfo  {journal} {Ann. Phys. (N.Y.)}\
  }\textbf {\bibinfo {volume} {334}},\ \bibinfo {pages} {321} (\bibinfo {year}
  {2013})}\BibitemShut {NoStop}%
\bibitem [{\citenamefont {Duan}\ and\ \citenamefont
  {Guo}(1997)}]{duan:ChinPhysLett:1997}%
  \BibitemOpen
  \bibfield  {author} {\bibinfo {author} {\bibfnamefont {L.-M.}\ \bibnamefont
  {Duan}}\ and\ \bibinfo {author} {\bibfnamefont {G.-C.}\ \bibnamefont {Guo}},\
  }\bibfield  {title} {\bibinfo {title} {Scheme for reducing decoherence in
  quantum computer memory by transformation to the coherence-preserving
  states},\ }\href@noop {} {\bibfield  {journal} {\bibinfo  {journal} {Chinese
  Phys. Lett.}\ }\textbf {\bibinfo {volume} {14}},\ \bibinfo {pages} {488}
  (\bibinfo {year} {1997})}\BibitemShut {NoStop}%
\bibitem [{\citenamefont {Duan}\ and\ \citenamefont {Guo}(1998)}]{duanguo2}%
  \BibitemOpen
  \bibfield  {author} {\bibinfo {author} {\bibfnamefont {L.~M.}\ \bibnamefont
  {Duan}}\ and\ \bibinfo {author} {\bibfnamefont {G.~C.}\ \bibnamefont {Guo}},\
  }\bibfield  {title} {\bibinfo {title} {Reducing decoherence in
  quantum-computer memory with all quantum bits coupling to the same
  environment},\ }\href@noop {} {\bibfield  {journal} {\bibinfo  {journal}
  {Phys. Rev. A}\ }\textbf {\bibinfo {volume} {57}},\ \bibinfo {pages} {737}
  (\bibinfo {year} {1998})}\BibitemShut {NoStop}%
\bibitem [{\citenamefont {Duan}\ and\ \citenamefont {Guo}(1999)}]{duanguo4}%
  \BibitemOpen
  \bibfield  {author} {\bibinfo {author} {\bibfnamefont {L.~M.}\ \bibnamefont
  {Duan}}\ and\ \bibinfo {author} {\bibfnamefont {G.~C.}\ \bibnamefont {Guo}},\
  }\bibfield  {title} {\bibinfo {title} {Quantum error correction with
  spatially correlated decoherence},\ }\href@noop {} {\bibfield  {journal}
  {\bibinfo  {journal} {Phys. Rev. A}\ }\textbf {\bibinfo {volume} {59}},\
  \bibinfo {pages} {4058} (\bibinfo {year} {1999})}\BibitemShut {NoStop}%
\bibitem [{\citenamefont {Shor}(1995)}]{shor}%
  \BibitemOpen
  \bibfield  {author} {\bibinfo {author} {\bibfnamefont {P.~W.}\ \bibnamefont
  {Shor}},\ }\bibfield  {title} {\bibinfo {title} {Scheme for reducing
  decoherence in quantum computer memory},\ }\href@noop {} {\bibfield
  {journal} {\bibinfo  {journal} {Phys. Rev. A}\ }\textbf {\bibinfo {volume}
  {52}},\ \bibinfo {pages} {R2493} (\bibinfo {year} {1995})}\BibitemShut
  {NoStop}%
\bibitem [{\citenamefont {Viola}\ and\ \citenamefont
  {Lloyd}(1998)}]{violalloyd}%
  \BibitemOpen
  \bibfield  {author} {\bibinfo {author} {\bibfnamefont {L.}~\bibnamefont
  {Viola}}\ and\ \bibinfo {author} {\bibfnamefont {S.}~\bibnamefont {Lloyd}},\
  }\bibfield  {title} {\bibinfo {title} {Dynamical suppression of decoherence
  in two-state quantum systems},\ }\href@noop {} {\bibfield  {journal}
  {\bibinfo  {journal} {Phys. Rev. A}\ }\textbf {\bibinfo {volume} {58}},\
  \bibinfo {pages} {2733} (\bibinfo {year} {1998})}\BibitemShut {NoStop}%
\bibitem [{\citenamefont {Novais}\ and\ \citenamefont
  {Baranger}(2006)}]{novais:PRL2006}%
  \BibitemOpen
  \bibfield  {author} {\bibinfo {author} {\bibfnamefont {E.}~\bibnamefont
  {Novais}}\ and\ \bibinfo {author} {\bibfnamefont {H.~U.}\ \bibnamefont
  {Baranger}},\ }\bibfield  {title} {\bibinfo {title} {Decoherence by
  correlated noise and quantum error correction},\ }\href@noop {} {\bibfield
  {journal} {\bibinfo  {journal} {Phys. Rev. Lett.}\ }\textbf {\bibinfo
  {volume} {97}},\ \bibinfo {pages} {040501} (\bibinfo {year}
  {2006})}\BibitemShut {NoStop}%
\bibitem [{\citenamefont {Alicki}(2006)}]{alicki}%
  \BibitemOpen
  \bibfield  {author} {\bibinfo {author} {\bibfnamefont {R.}~\bibnamefont
  {Alicki}},\ }\bibfield  {title} {\bibinfo {title} {A unified picture of
  decoherence control},\ }\href@noop {} {\bibfield  {journal} {\bibinfo
  {journal} {Chem. Phys.}\ }\textbf {\bibinfo {volume} {322}},\ \bibinfo
  {pages} {75} (\bibinfo {year} {2006})}\BibitemShut {NoStop}%
\bibitem [{\citenamefont {Lei}\ and\ \citenamefont
  {Zhang}(2011)}]{lei:PRA:2011}%
  \BibitemOpen
  \bibfield  {author} {\bibinfo {author} {\bibfnamefont {C.~U.}\ \bibnamefont
  {Lei}}\ and\ \bibinfo {author} {\bibfnamefont {W.-M.}\ \bibnamefont
  {Zhang}},\ }\bibfield  {title} {\bibinfo {title} {Decoherence suppression of
  open quantum systems through a strong coupling to non-Markovian
  reservoirs},\ }\href@noop {} {\bibfield  {journal} {\bibinfo  {journal}
  {Phys. Rev. A}\ }\textbf {\bibinfo {volume} {84}},\ \bibinfo {pages}
  {052116} (\bibinfo {year} {2011})}\BibitemShut {NoStop}%
\bibitem [{\citenamefont {Kim}\ \emph {et~al.}(2012)\citenamefont {Kim},
  \citenamefont {Lee}, \citenamefont {Kwon},\ and\ \citenamefont
  {Kim}}]{kimlee}%
  \BibitemOpen
  \bibfield  {author} {\bibinfo {author} {\bibfnamefont {Y.~S.}\ \bibnamefont
  {Kim}}, \bibinfo {author} {\bibfnamefont {J.~C.}\ \bibnamefont {Lee}},
  \bibinfo {author} {\bibfnamefont {O.}~\bibnamefont {Kwon}},\ and\ \bibinfo
  {author} {\bibfnamefont {Y.~H.}\ \bibnamefont {Kim}},\ }\bibfield  {title}
  {\bibinfo {title} {Protecting entanglement from decoherence using weak
  measurement and quantum measurement reversal},\ }\href@noop {} {\bibfield
  {journal} {\bibinfo  {journal} {Nature Phys.}\ }\textbf {\bibinfo {volume}
  {8}},\ \bibinfo {pages} {117} (\bibinfo {year} {2012})}\BibitemShut {NoStop}%
\bibitem [{\citenamefont {Bi}(2015)}]{bi}%
  \BibitemOpen
  \bibfield  {author} {\bibinfo {author} {\bibfnamefont {Q.}~\bibnamefont
  {Bi}},\ }\bibfield  {title} {\bibinfo {title} {Quantum computation in
  triangular decoherence-free subdynamic space},\ }\href@noop {} {\bibfield
  {journal} {\bibinfo  {journal} {Front. Phys.}\ }\textbf {\bibinfo {volume}
  {10}},\ \bibinfo {pages} {198} (\bibinfo {year} {2015})}\BibitemShut
  {NoStop}%
\bibitem [{\citenamefont {Ahsan}\ and\ \citenamefont
  {Naqvi}(2018)}]{ahsan:QIC:2018}%
  \BibitemOpen
  \bibfield  {author} {\bibinfo {author} {\bibfnamefont {M.}~\bibnamefont
  {Ahsan}}\ and\ \bibinfo {author} {\bibfnamefont {S.~A.~Z.}\ \bibnamefont
  {Naqvi}},\ }\bibfield  {title} {\bibinfo {title} {Performance of topological
  quantum error correction in the presence of correlated noise},\ }\href@noop
  {} {\bibfield  {journal} {\bibinfo  {journal} {Quantum Inf. Comput.}\
  }\textbf {\bibinfo {volume} {18}},\ \bibinfo {pages} {743} (\bibinfo {year}
  {2018})}\BibitemShut {NoStop}%
\bibitem [{\citenamefont {Ralph}\ and\ \citenamefont
  {Pienaar}(2014)}]{ralphpienaar}%
  \BibitemOpen
  \bibfield  {author} {\bibinfo {author} {\bibfnamefont {T.~C.}\ \bibnamefont
  {Ralph}}\ and\ \bibinfo {author} {\bibfnamefont {J.}~\bibnamefont
  {Pienaar}},\ }\bibfield  {title} {\bibinfo {title} {Entanglement decoherence
  in a gravitational well according to the event formalism},\ }\href@noop {}
  {\bibfield  {journal} {\bibinfo  {journal} {New J. Phys.}\ }\textbf {\bibinfo
  {volume} {16}},\ \bibinfo {pages} {085008} (\bibinfo {year} {2014})}\BibitemShut {NoStop}%
\bibitem [{\citenamefont {Beierle}\ \emph {et~al.}(2018)\citenamefont
  {Beierle}, \citenamefont {Zhang},\ and\ \citenamefont
  {Batelaan}}]{batelaan:NJP:2018}%
  \BibitemOpen
  \bibfield  {author} {\bibinfo {author} {\bibfnamefont {P.~J.}\ \bibnamefont
  {Beierle}}, \bibinfo {author} {\bibfnamefont {L.}~\bibnamefont {Zhang}},\
  and\ \bibinfo {author} {\bibfnamefont {H.}~\bibnamefont {Batelaan}},\
  }\bibfield  {title} {\bibinfo {title} {Experimental test of decoherence
  theory using electron matter waves},\ }\href@noop {} {\bibfield  {journal}
  {\bibinfo  {journal} {New J. Phys.}\ }\textbf {\bibinfo {volume} {20}},\
  \bibinfo {pages} {113030} (\bibinfo {year} {2018})}\BibitemShut {NoStop}%
\bibitem [{\citenamefont {Joshi}\ \emph {et~al.}(2018)\citenamefont {Joshi},
  \citenamefont {Pienaar}, \citenamefont {Ralph}, \citenamefont {Cacciapuoti},
  \citenamefont {McCutcheon}, \citenamefont {Rarity}, \citenamefont
  {Giggenbach}, \citenamefont {Lim}, \citenamefont {Makarov}, \citenamefont
  {Fuentes}, \citenamefont {Scheidl}, \citenamefont {Beckert}, \citenamefont
  {Bourennane}, \citenamefont {Bruschi}, \citenamefont {Cabello}, \citenamefont
  {Capmany}, \citenamefont {Carrasco-Casado}, \citenamefont {Diamanti},
  \citenamefont {Du\v{s}ek}, \citenamefont {Elser}, \citenamefont {Gulinatti},
  \citenamefont {Hadfield}, \citenamefont {Jennewein}, \citenamefont
  {Kaltenbaek}, \citenamefont {Krainak}, \citenamefont {Lo}, \citenamefont
  {Marquardt}, \citenamefont {Milburn}, \citenamefont {Peev}, \citenamefont
  {Poppe}, \citenamefont {Pruneri}, \citenamefont {Renner}, \citenamefont
  {Salomon}, \citenamefont {Skaar}, \citenamefont {Solomos}, \citenamefont
  {Stip\v{c}evic}, \citenamefont {Torres}, \citenamefont {Toyoshima},
  \citenamefont {Villoresi}, \citenamefont {Walrnsley}, \citenamefont {Weihs},
  \citenamefont {Weinfurter}, \citenamefont {Zeilinger}, \citenamefont
  {Zukowski},\ and\ \citenamefont {Ursin}}]{zeilinger:NJP:2018}%
  \BibitemOpen
  \bibfield  {author} {\bibinfo {author} {\bibfnamefont {S.~K.}\ \bibnamefont
  {Joshi}}, \bibinfo {author} {\bibfnamefont {J.}~\bibnamefont {Pienaar}},
  \bibinfo {author} {\bibfnamefont {T.~C.}\ \bibnamefont {Ralph}}, \bibinfo
  {author} {\bibfnamefont {L.}~\bibnamefont {Cacciapuoti}}, \bibinfo {author}
  {\bibfnamefont {W.}~\bibnamefont {McCutcheon}}, \bibinfo {author}
  {\bibfnamefont {J.}~\bibnamefont {Rarity}}, \bibinfo {author} {\bibfnamefont
  {D.}~\bibnamefont {Giggenbach}}, \bibinfo {author} {\bibfnamefont {J.~G.}\
  \bibnamefont {Lim}}, \bibinfo {author} {\bibfnamefont {V.}~\bibnamefont
  {Makarov}}, \bibinfo {author} {\bibfnamefont {I.}~\bibnamefont {Fuentes}},
  \bibinfo {author} {\bibfnamefont {T.}~\bibnamefont {Scheidl}}, \bibinfo
  {author} {\bibfnamefont {E.}~\bibnamefont {Beckert}}, \bibinfo {author}
  {\bibfnamefont {M.}~\bibnamefont {Bourennane}}, \bibinfo {author}
  {\bibfnamefont {D.~E.}\ \bibnamefont {Bruschi}}, \bibinfo {author}
  {\bibfnamefont {A.}~\bibnamefont {Cabello}}, \bibinfo {author} {\bibfnamefont
  {J.}~\bibnamefont {Capmany}}, \bibinfo {author} {\bibfnamefont
  {A.}~\bibnamefont {Carrasco-Casado}}, \bibinfo {author} {\bibfnamefont
  {E.}~\bibnamefont {Diamanti}}, \bibinfo {author} {\bibfnamefont
  {M.}~\bibnamefont {Du\v{s}ek}}, \bibinfo {author} {\bibfnamefont
  {D.}~\bibnamefont {Elser}}, \bibinfo {author} {\bibfnamefont
  {A.}~\bibnamefont {Gulinatti}}, \bibinfo {author} {\bibfnamefont {R.~H.}\
  \bibnamefont {Hadfield}}, \bibinfo {author} {\bibfnamefont {T.}~\bibnamefont
  {Jennewein}}, \bibinfo {author} {\bibfnamefont {R.}~\bibnamefont
  {Kaltenbaek}}, \bibinfo {author} {\bibfnamefont {M.~A.}\ \bibnamefont
  {Krainak}}, \bibinfo {author} {\bibfnamefont {H.~K.}\ \bibnamefont {Lo}},
  \bibinfo {author} {\bibfnamefont {C.}~\bibnamefont {Marquardt}}, \bibinfo
  {author} {\bibfnamefont {G.}~\bibnamefont {Milburn}}, \bibinfo {author}
  {\bibfnamefont {M.}~\bibnamefont {Peev}}, \bibinfo {author} {\bibfnamefont
  {A.}~\bibnamefont {Poppe}}, \bibinfo {author} {\bibfnamefont
  {V.}~\bibnamefont {Pruneri}}, \bibinfo {author} {\bibfnamefont
  {R.}~\bibnamefont {Renner}}, \bibinfo {author} {\bibfnamefont
  {C.}~\bibnamefont {Salomon}}, \bibinfo {author} {\bibfnamefont
  {J.}~\bibnamefont {Skaar}}, \bibinfo {author} {\bibfnamefont
  {N.}~\bibnamefont {Solomos}}, \bibinfo {author} {\bibfnamefont
  {M.}~\bibnamefont {Stip\v{c}evic}}, \bibinfo {author} {\bibfnamefont {J.~P.}\
  \bibnamefont {Torres}}, \bibinfo {author} {\bibfnamefont {M.}~\bibnamefont
  {Toyoshima}}, \bibinfo {author} {\bibfnamefont {P.}~\bibnamefont
  {Villoresi}}, \bibinfo {author} {\bibfnamefont {I.}~\bibnamefont
  {Walrnsley}}, \bibinfo {author} {\bibfnamefont {G.}~\bibnamefont {Weihs}},
  \bibinfo {author} {\bibfnamefont {H.}~\bibnamefont {Weinfurter}}, \bibinfo
  {author} {\bibfnamefont {A.}~\bibnamefont {Zeilinger}}, \bibinfo {author}
  {\bibfnamefont {M.}~\bibnamefont {Zukowski}},\ and\ \bibinfo {author}
  {\bibfnamefont {R.}~\bibnamefont {Ursin}},\ }\bibfield  {title} {\bibinfo
  {title} {Space QUEST mission proposal: Experimentally testing decoherence due
  to gravity},\ }\href@noop {} {\bibfield  {journal} {\bibinfo  {journal} {New
  J. Phys.}\ }\textbf {\bibinfo {volume} {20}},\ \bibinfo {pages} {063016}
  (\bibinfo {year} {2018})}\BibitemShut {NoStop}%
\bibitem [{\citenamefont {Xu}\ \emph {et~al.}(2019)\citenamefont {Xu},
  \citenamefont {Ma}, \citenamefont {Ren}, \citenamefont {Yong}, \citenamefont
  {Ralph}, \citenamefont {Liao}, \citenamefont {Yin}, \citenamefont {Liu},
  \citenamefont {Cai}, \citenamefont {Han}, \citenamefont {Wu}, \citenamefont
  {Wang}, \citenamefont {Li}, \citenamefont {Yang}, \citenamefont {Lin},
  \citenamefont {Li}, \citenamefont {Liu}, \citenamefont {Chen}, \citenamefont
  {Lu}, \citenamefont {Chen}, \citenamefont {Fan}, \citenamefont {Peng},\ and\
  \citenamefont {Pan}}]{xu}%
  \BibitemOpen
  \bibfield  {author} {\bibinfo {author} {\bibfnamefont {P.}~\bibnamefont
  {Xu}}, \bibinfo {author} {\bibfnamefont {Y.~Q.}\ \bibnamefont {Ma}}, \bibinfo
  {author} {\bibfnamefont {J.~G.}\ \bibnamefont {Ren}}, \bibinfo {author}
  {\bibfnamefont {H.~L.}\ \bibnamefont {Yong}}, \bibinfo {author}
  {\bibfnamefont {T.~C.}\ \bibnamefont {Ralph}}, \bibinfo {author}
  {\bibfnamefont {S.~K.}\ \bibnamefont {Liao}}, \bibinfo {author}
  {\bibfnamefont {J.}~\bibnamefont {Yin}}, \bibinfo {author} {\bibfnamefont
  {W.~Y.}\ \bibnamefont {Liu}}, \bibinfo {author} {\bibfnamefont {W.~Q.}\
  \bibnamefont {Cai}}, \bibinfo {author} {\bibfnamefont {X.}~\bibnamefont
  {Han}}, \bibinfo {author} {\bibfnamefont {H.~N.}\ \bibnamefont {Wu}},
  \bibinfo {author} {\bibfnamefont {W.~Y.}\ \bibnamefont {Wang}}, \bibinfo
  {author} {\bibfnamefont {F.~Z.}\ \bibnamefont {Li}}, \bibinfo {author}
  {\bibfnamefont {M.}~\bibnamefont {Yang}}, \bibinfo {author} {\bibfnamefont
  {F.~L.}\ \bibnamefont {Lin}}, \bibinfo {author} {\bibfnamefont
  {L.}~\bibnamefont {Li}}, \bibinfo {author} {\bibfnamefont {N.~L.}\
  \bibnamefont {Liu}}, \bibinfo {author} {\bibfnamefont {Y.~A.}\ \bibnamefont
  {Chen}}, \bibinfo {author} {\bibfnamefont {C.~Y.}\ \bibnamefont {Lu}},
  \bibinfo {author} {\bibfnamefont {Y.~B.}\ \bibnamefont {Chen}}, \bibinfo
  {author} {\bibfnamefont {J.~Y.}\ \bibnamefont {Fan}}, \bibinfo {author}
  {\bibfnamefont {C.~Z.}\ \bibnamefont {Peng}},\ and\ \bibinfo {author}
  {\bibfnamefont {J.~W.}\ \bibnamefont {Pan}},\ }\bibfield  {title} {\bibinfo
  {title} {Satellite testing of a gravitationally induced quantum decoherence
  model},\ }\href@noop {} {\bibfield  {journal} {\bibinfo  {journal} {Science}\
  }\textbf {\bibinfo {volume} {366}},\ \bibinfo {pages} {132} (\bibinfo {year}
  {2019})}\BibitemShut {NoStop}%
\bibitem [{\citenamefont {Kaplan}\ \emph {et~al.}(1998)\citenamefont {Kaplan},
  \citenamefont {Stifter}, \citenamefont {van Leeuwen}, \citenamefont
  {W.~E.~Lamb},\ and\ \citenamefont {Schleich}}]{kaplan:PhysScr:1998}%
  \BibitemOpen
  \bibfield  {author} {\bibinfo {author} {\bibfnamefont {A.~E.}\ \bibnamefont
  {Kaplan}}, \bibinfo {author} {\bibfnamefont {P.}~\bibnamefont {Stifter}},
  \bibinfo {author} {\bibfnamefont {K.~A.~H.}\ \bibnamefont {van Leeuwen}},
  \bibinfo {author} {\bibnamefont {W.~E.~Lamb,~Jr.}},\ and\
  \bibinfo {author} {\bibfnamefont {W.~P.}\ \bibnamefont {Schleich}},\
  }\bibfield  {title} {\bibinfo {title} {Intermode traces -- Fundamental
  interference phenomenon in quantum and wave physics},\ }\href@noop {}
  {\bibfield  {journal} {\bibinfo  {journal} {Phys. Scr.}\ }\textbf {\bibinfo
  {volume} {T76}},\ \bibinfo {pages} {93} (\bibinfo {year} {1998})}\BibitemShut
  {NoStop}%
\bibitem [{\citenamefont {Marzoli}\ \emph {et~al.}(1998)\citenamefont
  {Marzoli}, \citenamefont {Saif}, \citenamefont {Bialynicki-Birula},
  \citenamefont {Friesch}, \citenamefont {Kaplan},\ and\ \citenamefont
  {Schleich}}]{marzoli:ActaPhysSlo:1998}%
  \BibitemOpen
  \bibfield  {author} {\bibinfo {author} {\bibfnamefont {I.}~\bibnamefont
  {Marzoli}}, \bibinfo {author} {\bibfnamefont {F.}~\bibnamefont {Saif}},
  \bibinfo {author} {\bibfnamefont {I.}~\bibnamefont {Bialynicki-Birula}},
  \bibinfo {author} {\bibfnamefont {O.~M.}\ \bibnamefont {Friesch}}, \bibinfo
  {author} {\bibfnamefont {A.~E.}\ \bibnamefont {Kaplan}},\ and\ \bibinfo
  {author} {\bibfnamefont {W.~P.}\ \bibnamefont {Schleich}},\ }\bibfield
  {title} {\bibinfo {title} {Quantum carpets made simple},\ }\href@noop {}
  {\bibfield  {journal} {\bibinfo  {journal} {Acta Phys. Slovaca}\ }\textbf
  {\bibinfo {volume} {48}},\ \bibinfo {pages} {323} (\bibinfo {year}
  {1998})}\BibitemShut {NoStop}%
\bibitem [{\citenamefont {Kaplan}\ \emph {et~al.}(2000)\citenamefont {Kaplan},
  \citenamefont {Marzoli}, \citenamefont {W.~E.~Lamb},\ and\ \citenamefont
  {Schleich}}]{kaplan:PRA:2000}%
  \BibitemOpen
  \bibfield  {author} {\bibinfo {author} {\bibfnamefont {A.~E.}\ \bibnamefont
  {Kaplan}}, \bibinfo {author} {\bibfnamefont {I.}~\bibnamefont {Marzoli}},
  \bibinfo {author} {\bibnamefont {W.~E.~Lamb,~Jr.}},\ and\
  \bibinfo {author} {\bibfnamefont {W.~P.}\ \bibnamefont {Schleich}},\
  }\bibfield  {title} {\bibinfo {title} {Multimode interference: Highly regular
  pattern formation in quantum wave-packet evolution},\ }\href@noop {}
  {\bibfield  {journal} {\bibinfo  {journal} {Phys. Rev. A}\ }\textbf {\bibinfo
  {volume} {61}},\ \bibinfo {pages} {032101} (\bibinfo {year}
  {2000})}\BibitemShut {NoStop}%
\bibitem [{\citenamefont {Berry}\ \emph {et~al.}(2001)\citenamefont {Berry},
  \citenamefont {Marzoli},\ and\ \citenamefont
  {Schleich}}]{berry:PhysWorld:2001}%
  \BibitemOpen
  \bibfield  {author} {\bibinfo {author} {\bibfnamefont {M.}~\bibnamefont
  {Berry}}, \bibinfo {author} {\bibfnamefont {I.}~\bibnamefont {Marzoli}},\
  and\ \bibinfo {author} {\bibfnamefont {W.}~\bibnamefont {Schleich}},\
  }\bibfield  {title} {\bibinfo {title} {Quantum carpets, carpets of light},\
  }\href@noop {} {\bibfield  {journal} {\bibinfo  {journal} {Phys. World}\
  }\textbf {\bibinfo {volume} {14(6)}},\ \bibinfo {pages} {39} (\bibinfo {year}
  {2001})}\BibitemShut {NoStop}%
\bibitem [{\citenamefont {Bonifacio}\ \emph {et~al.}(2000)\citenamefont
  {Bonifacio}, \citenamefont {Marzoli},\ and\ \citenamefont
  {Schleich}}]{bonifacio}%
  \BibitemOpen
  \bibfield  {author} {\bibinfo {author} {\bibfnamefont {R.}~\bibnamefont
  {Bonifacio}}, \bibinfo {author} {\bibfnamefont {I.}~\bibnamefont {Marzoli}},\
  and\ \bibinfo {author} {\bibfnamefont {W.~P.}\ \bibnamefont {Schleich}},\
  }\bibfield  {title} {\bibinfo {title} {Non-dissipative decoherence for
  quantum carpets},\ }\href@noop {} {\bibfield  {journal} {\bibinfo  {journal}
  {J. Mod. Optic.}\ }\textbf {\bibinfo {volume} {47}},\ \bibinfo {pages} {2891}
  (\bibinfo {year} {2000})}\BibitemShut {NoStop}%
\bibitem [{\citenamefont {Kazemi}\ \emph {et~al.}(2013)\citenamefont {Kazemi},
  \citenamefont {Chaturvedi}, \citenamefont {Marzoli}, \citenamefont
  {O'Conell},\ and\ \citenamefont {Schleich}}]{schleich:NJP:2013}%
  \BibitemOpen
  \bibfield  {author} {\bibinfo {author} {\bibfnamefont {P.}~\bibnamefont
  {Kazemi}}, \bibinfo {author} {\bibfnamefont {S.}~\bibnamefont {Chaturvedi}},
  \bibinfo {author} {\bibfnamefont {I.}~\bibnamefont {Marzoli}}, \bibinfo
  {author} {\bibfnamefont {R.~F.}\ \bibnamefont {O'Conell}},\ and\ \bibinfo
  {author} {\bibfnamefont {W.~P.}\ \bibnamefont {Schleich}},\ }\bibfield
  {title} {\bibinfo {title} {Quantum carpets: A tool to observe decoherence},\
  }\href@noop {} {\bibfield  {journal} {\bibinfo  {journal} {New J. Phys.}\
  }\textbf {\bibinfo {volume} {15}},\ \bibinfo {pages} {013052} (\bibinfo
  {year} {2013})}\BibitemShut {NoStop}%
\bibitem [{\citenamefont {Sanz}(2014)}]{sanz:CJC:2014}%
  \BibitemOpen
  \bibfield  {author} {\bibinfo {author} {\bibfnamefont {A.~S.}\ \bibnamefont
  {Sanz}},\ }\bibfield  {title} {\bibinfo {title} {Effective Markovian
  description of decoherence in bound systems},\ }\href@noop {} {\bibfield
  {journal} {\bibinfo  {journal} {Can. J. Chem.}\ }\textbf {\bibinfo {volume}
  {92}},\ \bibinfo {pages} {168} (\bibinfo {year} {2014})}\BibitemShut
  {NoStop}%
\bibitem [{\citenamefont {Sanz}\ and\ \citenamefont
  {Borondo}(2007)}]{sanz:EPJD:2007}%
  \BibitemOpen
  \bibfield  {author} {\bibinfo {author} {\bibfnamefont {A.~S.}\ \bibnamefont
  {Sanz}}\ and\ \bibinfo {author} {\bibfnamefont {F.}~\bibnamefont {Borondo}},\
  }\bibfield  {title} {\bibinfo {title} {A quantum trajectory description of
  decoherence},\ }\href@noop {} {\bibfield  {journal} {\bibinfo  {journal}
  {Eur. Phys. J. D}\ }\textbf {\bibinfo {volume} {44}},\ \bibinfo {pages} {319}
  (\bibinfo {year} {2007})}\BibitemShut {NoStop}%
\bibitem [{\citenamefont {Sanz}\ and\ \citenamefont
  {Borondo}(2009)}]{sanz:CPL:2009-2}%
  \BibitemOpen
  \bibfield  {author} {\bibinfo {author} {\bibfnamefont {A.~S.}\ \bibnamefont
  {Sanz}}\ and\ \bibinfo {author} {\bibfnamefont {F.}~\bibnamefont {Borondo}},\
  }\bibfield  {title} {\bibinfo {title} {Contextuality, decoherence and quantum
  trajectories},\ }\href@noop {} {\bibfield  {journal} {\bibinfo  {journal}
  {Chem. Phys. Lett.}\ }\textbf {\bibinfo {volume} {478}},\ \bibinfo {pages}
  {301} (\bibinfo {year} {2009})}\BibitemShut {NoStop}%
\bibitem [{\citenamefont {Luis}\ and\ \citenamefont
  {Sanz}(2015)}]{luis:AOP:2015}%
  \BibitemOpen
  \bibfield  {author} {\bibinfo {author} {\bibfnamefont {A.}~\bibnamefont
  {Luis}}\ and\ \bibinfo {author} {\bibfnamefont {A.~S.}\ \bibnamefont
  {Sanz}},\ }\bibfield  {title} {\bibinfo {title} {What dynamics can be
  expected for mixed states in two-slit experiments?},\ }\href@noop {}
  {\bibfield  {journal} {\bibinfo  {journal} {Ann. Phys. (N.Y.)}\ }\textbf {\bibinfo
  {volume} {357}},\ \bibinfo {pages} {95} (\bibinfo {year} {2015})}\BibitemShut
  {NoStop}%
\bibitem [{\citenamefont {Sanz}\ and\ \citenamefont
  {Miret-Art\'es}(2008)}]{sanz:JPA:2008}%
  \BibitemOpen
  \bibfield  {author} {\bibinfo {author} {\bibfnamefont {A.~S.}\ \bibnamefont
  {Sanz}}\ and\ \bibinfo {author} {\bibfnamefont {S.}~\bibnamefont
  {Miret-Art\'es}},\ }\bibfield  {title} {\bibinfo {title} {A trajectory-based
  understanding of quantum interference},\ }\href@noop {} {\bibfield  {journal}
  {\bibinfo  {journal} {J. Phys. A: Math. Theor.}\ }\textbf {\bibinfo {volume}
  {41}},\ \bibinfo {pages} {435303} (\bibinfo {year} {2008})}\BibitemShut
  {NoStop}%
\bibitem [{\citenamefont {Sanz}(2019)}]{sanz:FrontPhys:2019}%
  \BibitemOpen
  \bibfield  {author} {\bibinfo {author} {\bibfnamefont {A.~S.}\ \bibnamefont
  {Sanz}},\ }\bibfield  {title} {\bibinfo {title} {Bohm's approach to quantum
  mechanics: Alternative theory or practical picture?},\ }\href@noop {}
  {\bibfield  {journal} {\bibinfo  {journal} {Front. Phys.}\ }\textbf {\bibinfo
  {volume} {14}},\ \bibinfo {pages} {11301} (\bibinfo {year}
  {2019})}\BibitemShut {NoStop}%
\bibitem{newref1}
 W.~K. Burns and A.~F. Milton, An analytic solution for mode coupling in optical waveguide branches, IEEE J. Quantum Electron. {\bf QE-16}, 446 (1980).
\bibitem [{\citenamefont {Soldano}\ and\ \citenamefont
  {Pennings}(1995)}]{pennings:JLightTech:1995}%
  \BibitemOpen
  \bibfield  {author} {\bibinfo {author} {\bibfnamefont {L.~B.}\ \bibnamefont
  {Soldano}}\ and\ \bibinfo {author} {\bibfnamefont {E.~C.~M.}\ \bibnamefont
  {Pennings}},\ }\bibfield  {title} {\bibinfo {title} {Optical multi-model
  interference devices based on self-imaging: Principles and applications},\
  }\href@noop {} {\bibfield  {journal} {\bibinfo  {journal} {J. Light. Technol.}\
  }\textbf {\bibinfo {volume} {13}},\ \bibinfo {pages} {615} (\bibinfo {year}
  {1995})}\BibitemShut {NoStop}%
\bibitem{newref3}
 L.~A. Coldren, S.~W. Corzine, and M.~L. Ma\v{s}anovi\'c, {\it Diode Lasers and Photonic Integrated Circuits}, 2nd Ed.\ (Wiley, Hoboken, NJ, 2012).
\bibitem{newref4}
 E. Andersson, T. Calarco, R. Folman, M. Andersson, B. Hessmo, and J. Schmiedmayer, Multimode interferometer for guided matter waves, Phys. Rev. Lett. {\bf 88}, 100401 (2002).
\bibitem [{\citenamefont {Tounli}\ \emph {et~al.}(2019)\citenamefont {Tounli},
  \citenamefont {Alvarado},\ and\ \citenamefont {Sanz}}]{sanz:PhysScr:2019}%
  \BibitemOpen
  \bibfield  {author} {\bibinfo {author} {\bibfnamefont {J.}~\bibnamefont
  {Tounli}}, \bibinfo {author} {\bibfnamefont {A.}~\bibnamefont {Alvarado}},\
  and\ \bibinfo {author} {\bibfnamefont {A.~S.}\ \bibnamefont {Sanz}},\
  }\bibfield  {title} {\bibinfo {title} {Boundary bound diffraction: A combined
  spectral and bohmian mechanics},\ }\href@noop {} {\bibfield  {journal}
  {\bibinfo  {journal} {Phys. Scr.}\ }\textbf {\bibinfo {volume} {94}},\
  \bibinfo {pages} {035202} (\bibinfo {year} {2019})}\BibitemShut {NoStop}%
\bibitem{footnote}
 By referring to the localized state entering the cavity as the input signal, the
 connection to the language employed in fiber optic communications becomes closer,
 particularly in the realm of quantum optics with single-photon transmission,
 where this type of scenario could be experimentally tested.
\bibitem [{\citenamefont {Schiff}(1968)}]{schiff-bk}%
  \BibitemOpen
  \bibfield  {author} {\bibinfo {author} {\bibfnamefont {L.~I.}\ \bibnamefont
  {Schiff}},\ }\href@noop {} {\emph {\bibinfo {title} {Quantum Mechanics}}},\
  \bibinfo {edition} {3rd}\ Ed.\ (\bibinfo  {publisher} {McGraw-Hill},\
  \bibinfo {address} {Singapore},\ \bibinfo {year} {1968})\BibitemShut
  {NoStop}%
\bibitem [{\citenamefont {Holland}(1993)}]{holland-bk}%
  \BibitemOpen
  \bibfield  {author} {\bibinfo {author} {\bibfnamefont {P.~R.}\ \bibnamefont
  {Holland}},\ }\href@noop {} {\emph {\bibinfo {title} {The Quantum Theory of
  Motion}}}\ (\bibinfo  {publisher} {Cambridge University Press},\ \bibinfo
  {address} {Cambridge},\ \bibinfo {year} {1993})\BibitemShut {NoStop}%
\bibitem [{\citenamefont {Sanz}\ \emph {et~al.}(2012)\citenamefont {Sanz},
  \citenamefont {Campos-Mart{\'\i}nez},\ and\ \citenamefont
  {Miret-Art\'es}}]{sanz:JOSAA:2012}%
  \BibitemOpen
  \bibfield  {author} {\bibinfo {author} {\bibfnamefont {A.~S.}\ \bibnamefont
  {Sanz}}, \bibinfo {author} {\bibfnamefont {J.}~\bibnamefont
  {Campos-Mart{\'\i}nez}},\ and\ \bibinfo {author} {\bibfnamefont
  {S.}~\bibnamefont {Miret-Art\'es}},\ }\bibfield  {title} {\bibinfo {title}
  {Transmission properties in waveguides: An optical streamline analysis},\
  }\href@noop {} {\bibfield  {journal} {\bibinfo  {journal} {J. Opt. Am. Soc.
  A}\ }\textbf {\bibinfo {volume} {29}},\ \bibinfo {pages} {695} (\bibinfo
  {year} {2012})}\BibitemShut {NoStop}%
\bibitem [{\citenamefont {Sanz}\ and\ \citenamefont
  {Miret-Art\'es}(2012)}]{sanz-bk-1}%
  \BibitemOpen
  \bibfield  {author} {\bibinfo {author} {\bibfnamefont {A.~S.}\ \bibnamefont
  {Sanz}}\ and\ \bibinfo {author} {\bibfnamefont {S.}~\bibnamefont
  {Miret-Art\'es}},\ }\href@noop {} {\emph {\bibinfo {title} {A Trajectory
  Description of Quantum Processes. I. Fundamentals}}},\ \bibinfo {series}
  {Lecture Notes in Physics}, Vol.\ \bibinfo {volume} {850}\ (\bibinfo
  {publisher} {Springer},\ \bibinfo {address} {Berlin},\ \bibinfo {year}
  {2012})\BibitemShut {NoStop}%
\bibitem [{\citenamefont {Tounli}\ and\ \citenamefont
  {Sanz}(tion)}]{tounli:arxiv:2021}%
  \BibitemOpen
  \bibfield  {author} {\bibinfo {author} {\bibfnamefont {J.}~\bibnamefont
  {Tounli}}\ and\ \bibinfo {author} {\bibfnamefont {A.~S.}\ \bibnamefont
  {Sanz}},\ }\href@noop {} {\  (\bibinfo {year} {unpublished})}\BibitemShut
  {NoStop}%
\bibitem [{\citenamefont {Sanz}\ \emph {et~al.}(2002)\citenamefont {Sanz},
  \citenamefont {Borondo},\ and\ \citenamefont
  {Miret-Art\'es}}]{sanz:JPCM:2002}%
  \BibitemOpen
  \bibfield  {author} {\bibinfo {author} {\bibfnamefont {A.~S.}\ \bibnamefont
  {Sanz}}, \bibinfo {author} {\bibfnamefont {F.}~\bibnamefont {Borondo}},\ and\
  \bibinfo {author} {\bibfnamefont {S.}~\bibnamefont {Miret-Art\'es}},\
  }\bibfield  {title} {\bibinfo {title} {Particle diffraction studied using
  quantum trajectories},\ }
  {\bibfield  {journal} {\bibinfo  {journal} {J. Phys.: Condens. Matter}\
  }\textbf {\bibinfo {volume} {14}},\ \bibinfo {pages} {6109} (\bibinfo {year}
  {2002})}\BibitemShut {NoStop}%
\bibitem [{\citenamefont {Omn{\`e}s}(1992)}]{omnes:RMP:1992}%
  \BibitemOpen
  \bibfield  {author} {\bibinfo {author} {\bibfnamefont {R.}~\bibnamefont
  {Omn{\`e}s}},\ }\bibfield  {title} {\bibinfo {title} {Consistent
  interpretations of quantum mechanics},\ }\href@noop {} {\bibfield  {journal}
  {\bibinfo  {journal} {Rev. Mod. Phys.}\ }\textbf {\bibinfo {volume} {64}},\
  \bibinfo {pages} {339} (\bibinfo {year} {1992})}\BibitemShut {NoStop}%
\bibitem [{\citenamefont {Joos}(1996)}]{joos:bk:1996}%
  \BibitemOpen
  \bibfield  {author} {\bibinfo {author} {\bibfnamefont {E.}~\bibnamefont
  {Joos}},\ }\bibinfo {title} {Decoherence and the Appearance of a Classical
  World in Quantum Theory}\ (\bibinfo  {publisher} {Springer},\ \bibinfo
  {address} {Berlin},\ \bibinfo {year} {1996}),\ Chap.~3, pp.\ \bibinfo {pages}
  {41--180},\ \bibinfo {edition} {2nd}\ Ed.\BibitemShut {Stop}%
\bibitem [{\citenamefont {Elran}\ and\ \citenamefont
  {Brumer}(2004)}]{brumer-elran:JCP:2004}%
  \BibitemOpen
  \bibfield  {author} {\bibinfo {author} {\bibfnamefont {Y.}~\bibnamefont
  {Elran}}\ and\ \bibinfo {author} {\bibfnamefont {P.}~\bibnamefont {Brumer}},\
  }\bibfield  {title} {\bibinfo {title} {Decoherence in an anharmonic
  oscillator coupled to a thermal environment: A semiclassical forward-backward
  approach},\ }\href@noop {} {\bibfield  {journal} {\bibinfo  {journal} {J.
  Chem. Phys.}\ }\textbf {\bibinfo {volume} {121}},\ \bibinfo {pages} {2673}
  (\bibinfo {year} {2004})}\BibitemShut {NoStop}%
\bibitem [{\citenamefont {Elran}\ and\ \citenamefont
  {Brumer}(2013)}]{brumer-elran:JCP:2013}%
  \BibitemOpen
  \bibfield  {author} {\bibinfo {author} {\bibfnamefont {Y.}~\bibnamefont
  {Elran}}\ and\ \bibinfo {author} {\bibfnamefont {P.}~\bibnamefont {Brumer}},\
  }\bibfield  {title} {\bibinfo {title} {Quantum decoherence of I$_2$ in liquid
  xenon: A classical Wigner approach},\ }\href@noop {} {\bibfield  {journal}
  {\bibinfo  {journal} {J. Chem. Phys.}\ }\textbf {\bibinfo {volume} {138}},\
  \bibinfo {pages} {234308} (\bibinfo {year} {2013})}\BibitemShut {NoStop}%
\bibitem [{\citenamefont {Sanz}\ and\ \citenamefont
  {Miret-Art\'es}(2007)}]{sanz:JCP-Talbot:2007}%
  \BibitemOpen
  \bibfield  {author} {\bibinfo {author} {\bibfnamefont {A.~S.}\ \bibnamefont
  {Sanz}}\ and\ \bibinfo {author} {\bibfnamefont {S.}~\bibnamefont
  {Miret-Art\'es}},\ }\bibfield  {title} {\bibinfo {title} {A causal look into
  the quantum Talbot effect},\ }\href@noop {} {\bibfield  {journal} {\bibinfo
  {journal} {J. Chem. Phys.}\ }\textbf {\bibinfo {volume} {126}},\ \bibinfo
  {pages} {234106} (\bibinfo {year} {2007})}\BibitemShut {NoStop}%
\bibitem [{\citenamefont {Wang}\ \emph {et~al.}(2001)\citenamefont {Wang},
  \citenamefont {Thoss}, \citenamefont {Sorge}, \citenamefont {Gelabert},
  \citenamefont {Gim\'enez},\ and\ \citenamefont {Miller}}]{miller:JCP-1:2001}%
  \BibitemOpen
  \bibfield  {author} {\bibinfo {author} {\bibfnamefont {H.}~\bibnamefont
  {Wang}}, \bibinfo {author} {\bibfnamefont {M.}~\bibnamefont {Thoss}},
  \bibinfo {author} {\bibfnamefont {K.~L.}\ \bibnamefont {Sorge}}, \bibinfo
  {author} {\bibfnamefont {R.}~\bibnamefont {Gelabert}}, \bibinfo {author}
  {\bibfnamefont {X.}~\bibnamefont {Gim\'enez}},\ and\ \bibinfo {author}
  {\bibfnamefont {W.~H.}\ \bibnamefont {Miller}},\ }\bibfield  {title}
  {\bibinfo {title} {Semiclassical description of quantum coherence effects and
  their quenching: A forward-backward initial value representation study},\
  }\href@noop {} {\bibfield  {journal} {\bibinfo  {journal} {J. Chem. Phys.}\
  }\textbf {\bibinfo {volume} {114}},\ \bibinfo {pages} {2562} (\bibinfo {year}
  {2001})}\BibitemShut {NoStop}%
\end{thebibliography}

%

\end{document}